\documentclass[10pt,aps.prl,twocolumn,superscriptaddress]{revtex4-1} 
\usepackage{graphicx}
\usepackage{amsmath}
\usepackage{amssymb}
 \usepackage{wrapfig, framed, caption}
\usepackage{float}
\usepackage{subfig}
\usepackage{tabularx}
\usepackage[normalem]{ulem}
\usepackage{color, colortbl}
\usepackage{cancel}
\usepackage{epstopdf}

\newcommand{\enquote}[1]{``#1''}
\begin{document}
\title{Molecular structure and reversible photodegradation in anthraquinone dyes}
\author{Prabodh Dhakal}
\email{dprabodh@wsu.edu}
\author{Mark G. Kuzyk}
\email{mgk.wsu@gmail.com}
\affiliation{Department of Physics and Astronomy, Washington State University,
Pullman, WA 99164-2814}
\date{\today}

\begin{abstract}
Reversible photodegradation is a process that has been observed in several dye molecules, but the underlying mechanisms are not still well understood.  In this contribution, we characterize a series of anthraquinone dyes to determine how self-healing depends on molecular structure.  Past studies have used probing techniques that rely on linear absorption, two-photon fluorescence, and amplified spontaneous emission.  Each of these probes provide an indirect measure of the populations of the damaged and undamaged species, requiring calibrations or assumptions to be made that might affect the accuracy of the results.  The present studies use fluorescence as a probe, which is shown to directly measure the undamaged population.  It is found that certain anthraquinone classes share common structural features that are associated with self healing.  Furthermore, the time and temperature dependence of photodegradation and self-healing is found to be consistent with the domain model of self healing.

\vspace{1em}
OCIS Codes:

\end{abstract}

\maketitle

\section{Introduction}

Dye molecules under high-intensity illumination degrade,\cite{zhang98.01,galvan20.01} a process called photodegradation.  Photodegradation is typically characterized by the irreversible breaking of bonds, which is accompanied by a change in the linear absorption spectrum when the molecule transitions from its pristine state to a damaged one.  Dye lasers are a well known example of a liquid solution,\cite{fletc78.01,rahn97.01,kuzne85.01,mhibi13.01} solid solution\cite{popov.98.01,kuria02.01,mani13.01} or in a sol-gel\cite{deshp10.01,yyang13.01} in which the active dye solute degrades under such photochemical reactions upon prolonged exposure.

Self-healing of the optical properties of a dye-doped polymer fiber was first observed by Peng and coworkers with fluorescence probing,\cite{Peng98.01} but was not further studied.  Howell independently observed self-healing in the anthraquinone dye Disperse Orange 11 (DO11) doped into poly (methyl methacrylate) (PMMA) polymer using Amplified Spontaneous Emission (ASE) as a probe,\cite{howel02.01} and was the first to apply kinetic models to determine rate constants and predict the intensity-dependence.  These studies suggested that the decay rate decreases when cylcing the material through several periods of decay/recovery and such cycling could also increase the ASE efficiency.  These observations led to the proposal that cycling dye doped polymers through intervals of damage and recovery might be a method for making more robust materials.

It is important to stress that these self-healing materials are not specially designed to heal after photodegradation.  The materials studied here and the processes responsible are not related to the large body of literature on self-repair of polymers after microcracking, where the material incorporates a microencapsulated healing agent that is released upon crack intrusion\cite{white01.01}. In the present studies, the dopant dye chromophores and the surrounding polymer undergo light-induced chemical processes in which bonds may be broken or rearranged, which are reversed upon healing -- probably at the molecular level.

Howell also showed that dyes that irreversibly photodegrade in liquid solution\cite{howel04.01} self-heal when the same experimental protocol is used with dyes that are instead embedded in a polymer host.  In the dye laser literature, self healing is associated with the replenishment of fresh dye into regions where dye was damaged through mass transport and mixing.  Our definition of self healing is the literal recovery of the pristine molecule from its damaged state rather than a damaged molecule being merely replaced by a a fresh one.  Howell meticulously made a sample chamber that was uniformly illuminate to eliminate reservoirs of undamaged dyes.

Embaye et al. used photochromism experiments during decay to show that reorientational diffusion (sometimes called orientational hole burning) is absent, so is not a likely mechanism.\cite{embay07.01} Ramini and Dawson applied an analysis of samples that were imaged during decay and recovery to show that mass transport of molecules away from the burn area during damage and towards it in the dark during recovery was absent based on the evolution of the optically-imaged burn profile.\cite{ramin11.01}  These set of experiments eliminated all of the standard explanations and suggested that a new type of mechanism was at play.

Embaye's observations that the recovery rate accelerates for samples with high dye concentrations\cite{embay07.01} suggested that aggregates of dyes might be responsible for recovery, where larger aggregates both protect the molecules within them and foster their recovery.  The fact that increasing the temperature decreases the recovery rate led to the hypothesis that domains form in analogy to condensates, where the average size of an aggregate decreases as the temperature increases.

Ramini and coworkers developed a statistical mechanical condensation model based on only one adjustable parameter that quantifies the sticking energy of a molecule to a domain.\cite{ramin12.01}  This model was applied to Disperse Red 1 azo-dye-doped PMMA polymer data to determine the sticking energy (about $0.26 \pm 0.1 eV$), from which the distribution of sizes was determined. The measured temperature-dependence agreed with the model over a wide range of measurements.

Later, Anderson developed an imaging technique that has the ability to determine the population of damaged molecules as a function of intensity over a broad range of doses,\cite{Anderson11.01} which was used to fine tune the domain model.\cite{ramin13.01} The resulting model was found to apply to even a broader range of experimental parameters, strengthening the case for its veracity.

The domain model assumes that each molecule sticks to two others, so domains are inherently quasi-one-dimensional.  Domain models with higher-order dimensionality make predictions that disagree with the data.  To further investigate this property, the effect of an applied electric field on the kinetics of photodegradation and self-healing were experimentally studied.\cite{anders14.01} The quasi-one-dimensional model was generalized to take into account the effect of an applied field, which induces a dipole moment in each molecule, thus affecting the interactions between them.  These self-consistent dipole field models correctly predict the measurements, reinforcing the confidence in the model.

There is no doubt that self healing is a real phenomenon.  Indeed, the effect is being used to make lasers that heal.\cite{ander15.01, ander15.02}.  However, many questions remain, such as the mechanisms of domain formation and how domains mediate healing.  Dirk proposed that a TICT configuration might be responsible.\cite{dirk12.01} To test such hypotheses requires independent measurement techniques.  Each suffers from unique challenges.

Amplified spontaneous emission in ASE-active materials such as the anthraquinones studied in the present work is a highly sensitive probe because it differentiates strongly between the pristine molecule and the damaged species.  Extensive measurements of the dye disperse orange 11 (DO11) suggest that ASE emissions are registered only for the undamaged molecule and since the ASE intensity is a nonlinear function of the population density, its sensitivity is greater than for a linear measurements such as absorption spectroscopy.

The downside is that ASE is also a highly sensitive function of pump intensity, thus being more susceptible to laser drift.  As such, photodegradation and self-healing are easily observed; but, the measured time constants are prone to inconsistencies. Indeed, drift over the long time periods required to fully characterize recovery (on the order of 24 hours) can distort what is a simple single exponential curve to what appears to be multi-exponential.  Furthermore, calibrating the ASE intensity as a function of concentration requires carefully controlled measurements that take into account absorption gradients that develop during photodegradation.  These factors can lead to difficulties in obtaining quantitative data.

Linear absorbance, on the other hand, is a linear process, so the the amount of light absorbed is directly proportional to the number of molecules in the beam's path.  However, since the absorption spectrum of the pristine and damaged species overlap, both contribute.  To unentangle one from the other requires the determination of the wavelength dependence of the absorbtion cross section of the pristine dye (which is simple since pristine samples are available); but a purely decayed sample cannot be made because self healing continually converts the damaged product back to the original one.  Separating the two is especially difficult if the two spectra have substantial overlap.  The issue is complicated even more by the fact that an irreversible species is also present, so that when enough of the self-healing species is present to be measurable, the data is contaminated with the irreversible species.

In the present work, we use fluorescence spectroscopy as a probe.\cite{dhaka12.01}  It is a linear process so does not have to be well calibrated, and as we will later show, the damaged species do not appear to contribute a measurable amount of signal in the spectral range of the measurement.  As such, fluorescence probing appears to have all of the advantages without any of the drawbacks.  We describe when the technique is applicable and use it to characterize self healing in a large number of anthraquinone molecules to both characterize the kinetics and to determine which structural features are associated with healing.

\section{Experiment}

\begin{figure}
\begin{center}
\includegraphics{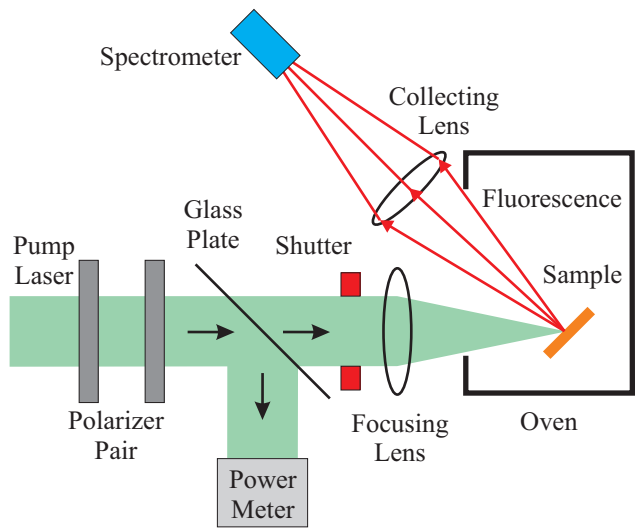}
\end{center}
\caption{Temperature-dependent fluorescence experiment.  The pump beam both damages the sample and produces fluorescence that is used to monitor the population of undamaged molecules.}
\label{fig:Experiment}
\end{figure}

\begin{table}[h]
\begin{framed}
\caption{Molecules from the {\em Amine Class} of anthraquinone molecules.  Dye C, with a diagonal red slash, is not observed to self-heal reproducibly, though some runs show recovery above noise levels.}\label{tab:AmineMolecules}
  \centering
  \begin{tabular}{c c c}
  \includegraphics[width = 1.1 in]{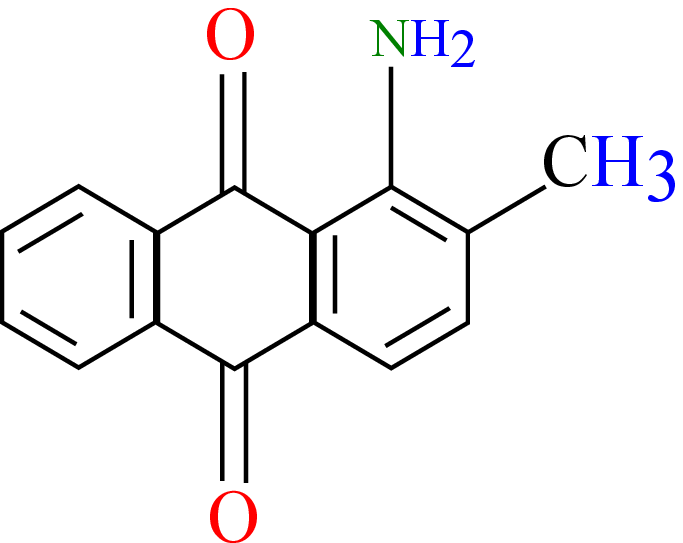} & & \includegraphics[width = 1 in]{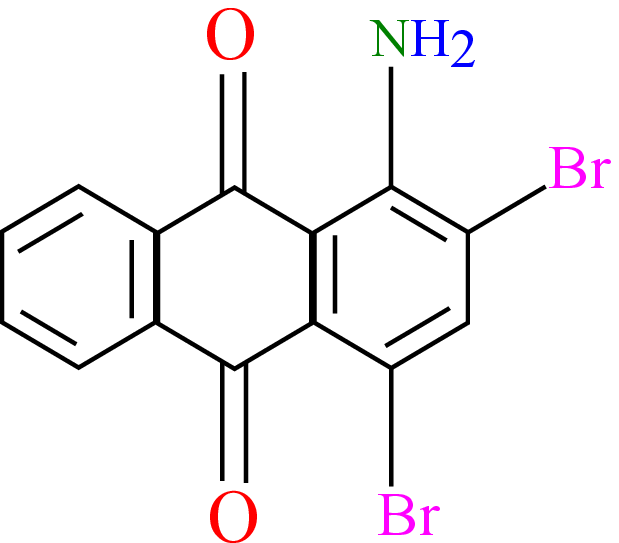} \\
  Dye A & & Dye B \\
   & & \\
  { \color{red} \cancel{\includegraphics[width = 0.8 in]{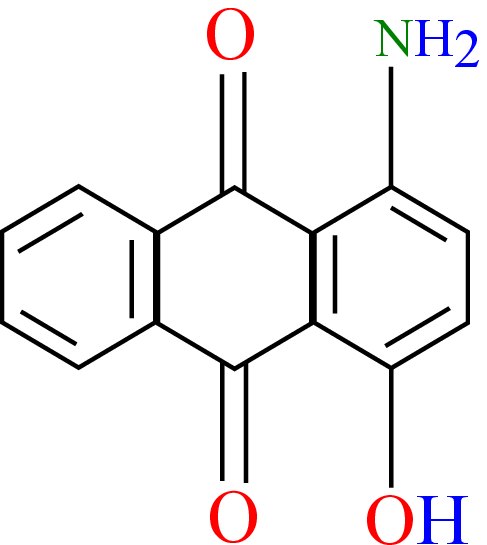}}} & & \includegraphics[width = 0.8 in]{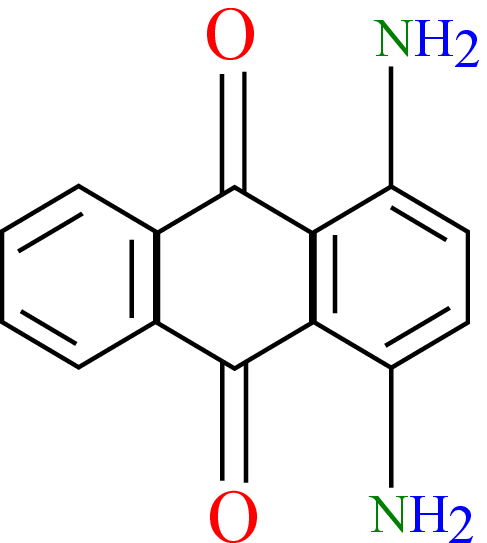}  \\
  Dye C & & Dye H \\
   & & \\
  \includegraphics[width = 0.8 in]{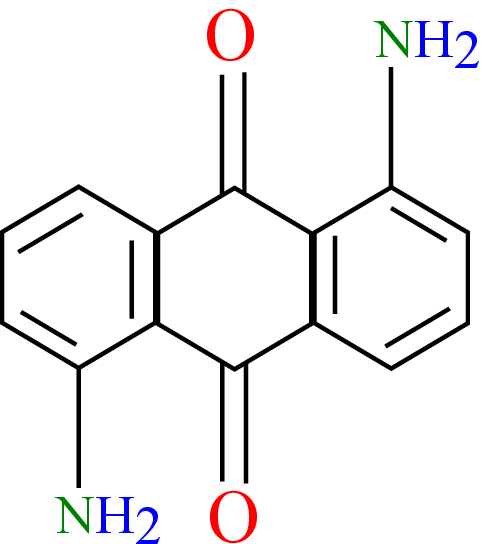} & & \includegraphics[width = 0.9 in]{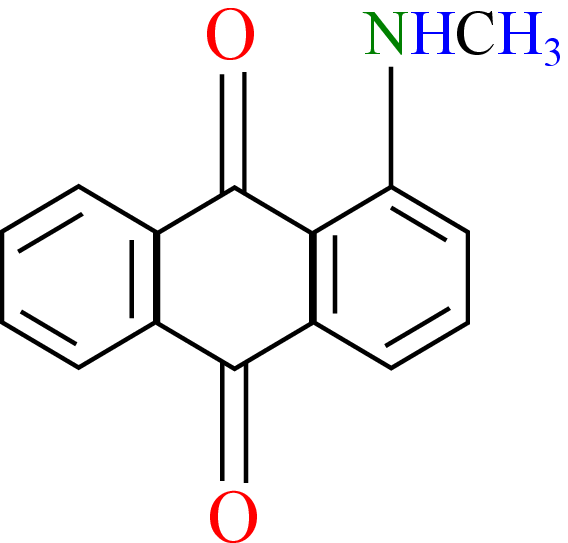} \\
  Dye I & & Dye P \\
\end{tabular}
\end{framed}
\end{table}

\begin{table}[h]
\begin{framed}
\caption{Molecules from the {\em Hydroxy Class}.  The molecule with a diagonal red slash does not self-heal.}\label{tab:HydroxyMolecules}
  \centering
  \begin{tabular}{c c c}
  \includegraphics[width = 1 in]{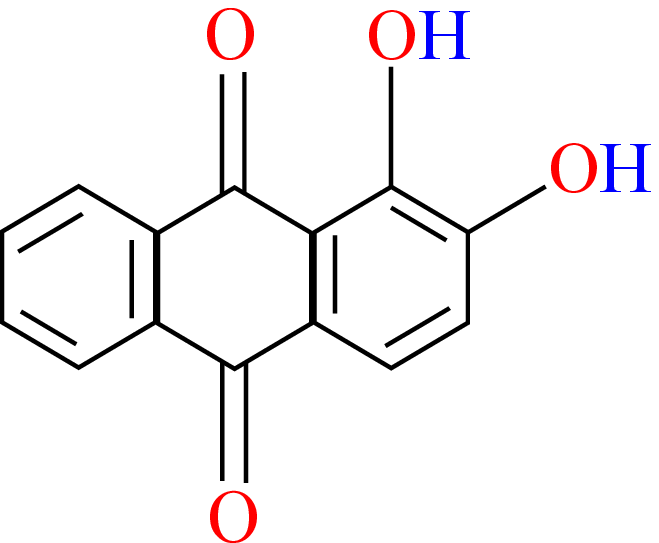} & &  \includegraphics[width = .8 in]{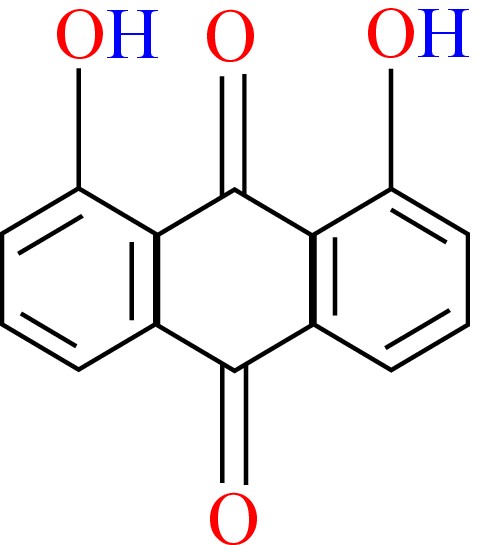} \\
  Dye J & & Dye K \\
   & & \\
  \includegraphics[width = .7 in]{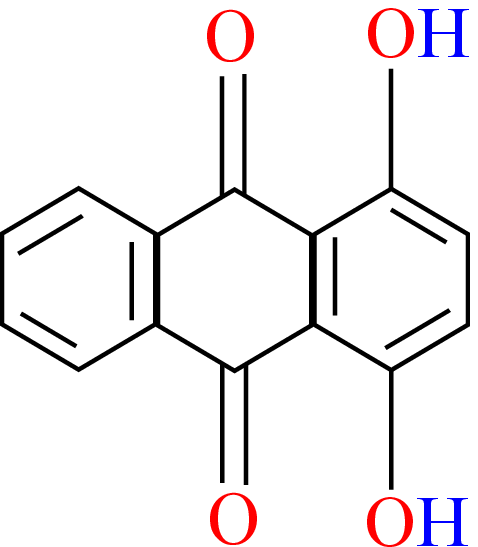} & & { \color{red} \cancel{\includegraphics[width = 1.1 in]{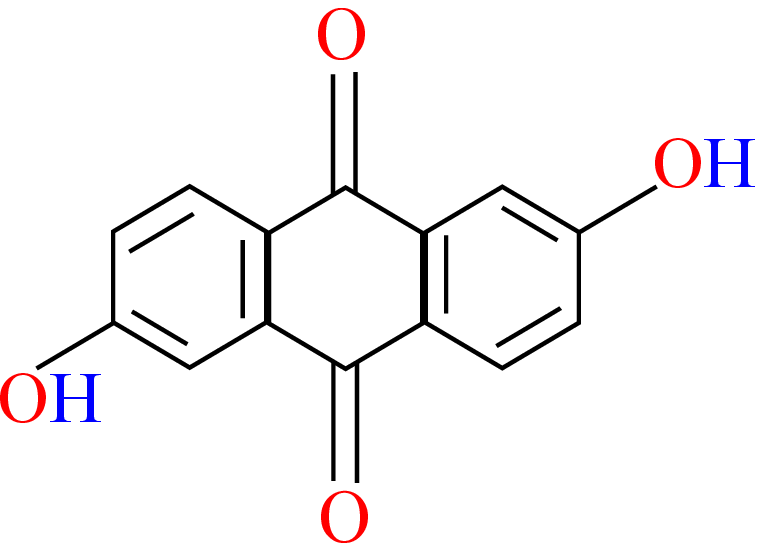}}} \\
  Dye L & & Dye M \\
\end{tabular}
\end{framed}
\end{table}

\begin{table}[h]
\begin{framed}
\caption{Molecules from the {\em Other Class}.  The molecules with a diagonal red slash do not self-heal.}\label{tab:OtherMolecules}
  \centering
  \begin{tabular}{c c c}
  { \color{red} \cancel{\includegraphics[width = .7 in]{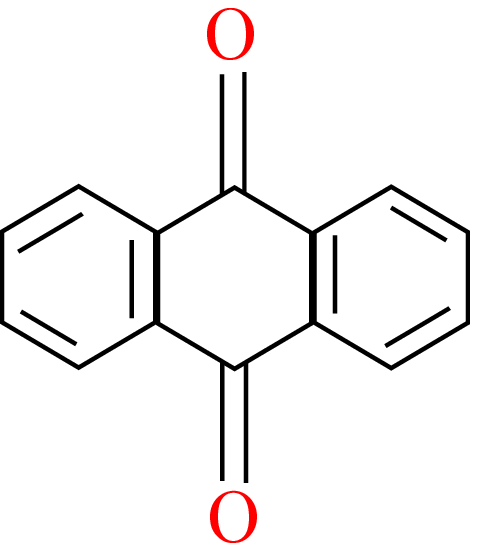}}} & & \includegraphics[width = 0.8 in]{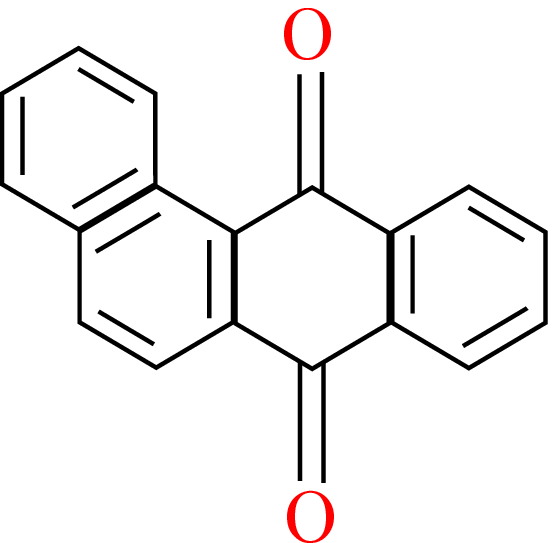} \\
  Dye D & & Dye E \\
     & & \\
  { \color{red} \cancel{\includegraphics[width = 1 in]{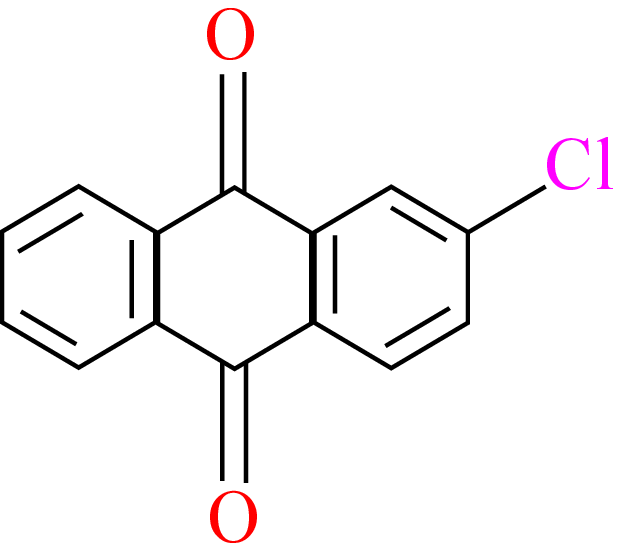}}} & & { \color{red} \cancel{\includegraphics[width = 0.8 in]{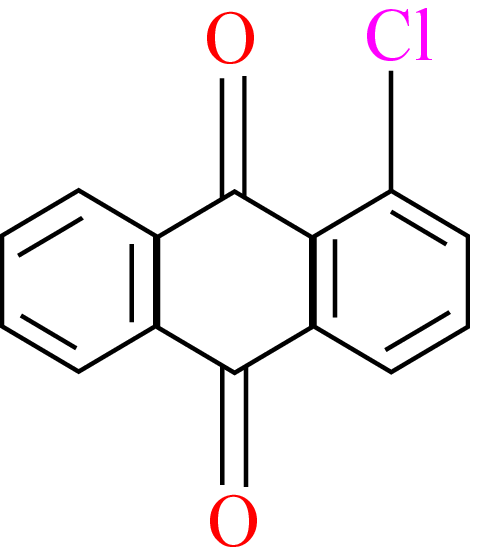}}} \\
  Dye F & & Dye G \\
   & & \\
  { \color{red} \cancel{\includegraphics[width = 1 in]{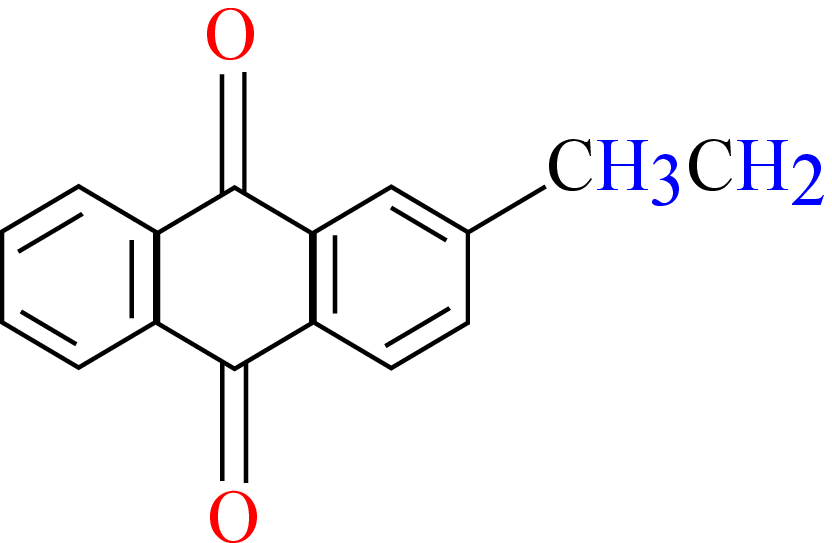}}} & &  { \color{red} \cancel{\includegraphics[width = 1 in]{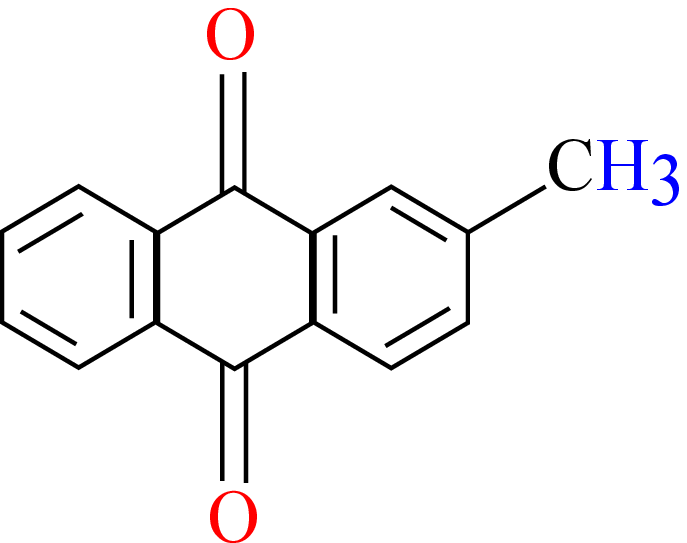}}} \\
  Dye N & & Dye O \\
\end{tabular}
\end{framed}
\end{table}

Figure \ref{fig:Experiment} shows a schematic diagram of the fluorescence spectroscopy experiment that is used to probe optical damage.  A pump laser from a Coherent Innova model 70C series ion laser at a wavelength of 488nm and $3.46 \times 10^5 W/m^2$ average passes a pair of polarizers, which are used to control the power.  A glass plate splits a small portion of the beam to a Thorlabs model S20MM power meter that monitors the stability of the beam, which is focused onto a thin-film dye-doped polymer sample in an oven chamber.  The temperature is controlled with a resistive heater driven by a Micromega CN 77322-C2 unit, and the temperature is monitored with an Omega model CN-2010 Thermocouple.  A shutter is used to turn the pump on and off as needed.  The fluorescence produced by the sample is collected with a lens that directs the light to a fiber which guides the light into an Ocean Optics model SD2000 spectrometer.

In a typical run, the fluorescence spectrum is measured periodically while the pump remains on to photo-degrade the sample.  The pump beam is then blocked with a shutter to allow the sample to self heal.  During self healing, the shutter is opened occasionally to probe recovery, but for short enough time intervals -- typically 12 to 15 seconds -- to prevent significant material damage.  A decay and recovery run is repeated several times on different parts of the sample to determine point-to-point variations.  The experiment is repeated over a range of temperatures to probe material photodegradation and self-healing in the time-temperature domain.

The linear absorption spectra of dye-doped films are measured in the usual way using a white light source and an Ocean optics spectrometer with appropriate corrections to take into account the substrate contribution and the spectrum of the light source.

Tables \ref{tab:AmineMolecules}, \ref{tab:HydroxyMolecules}, and \ref{tab:OtherMolecules} show the molecules studied in this work.  The molecules are divided into three classes: the Amine class contains an $NH_2$ group, the Hydroxy Class an $OH$ group, and the Other Class lacks $NH_2$ and $OH$, but has groups such as $Cl$, $CH_3$, $C_2H_5$ and $NHCH_3$.

\begin{figure*}
\includegraphics[width = 2.35 in]{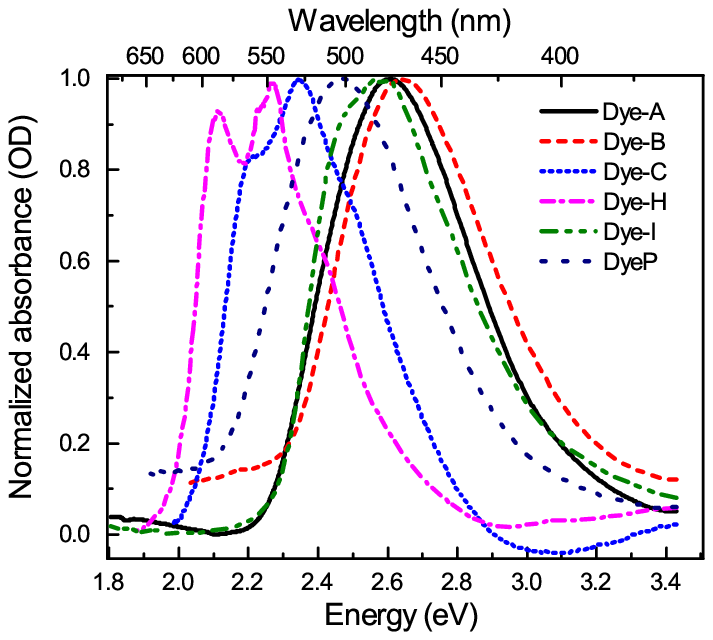}\includegraphics[width = 2.35 in]{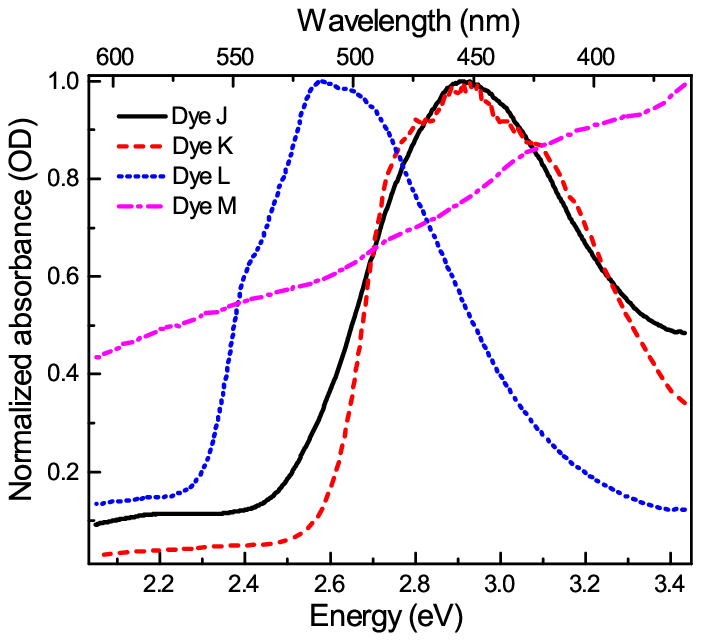}\includegraphics[width = 2.35 in]{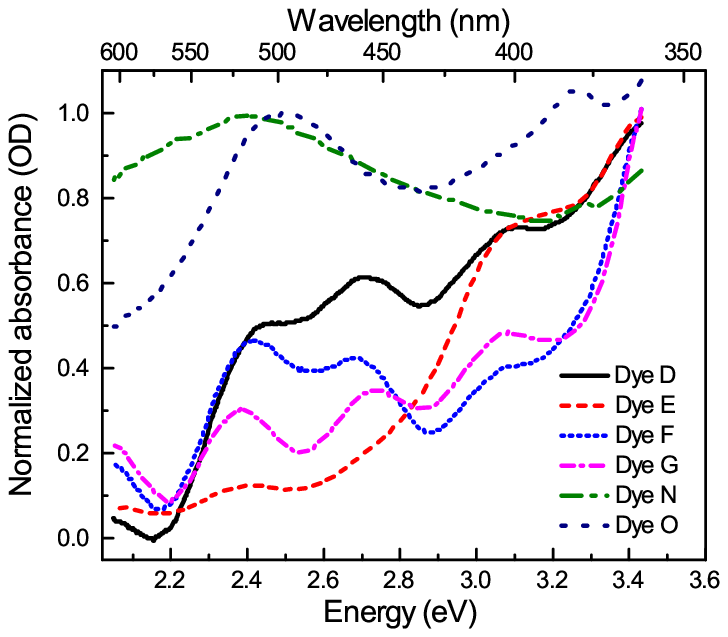}
\includegraphics[width = 2.35 in]{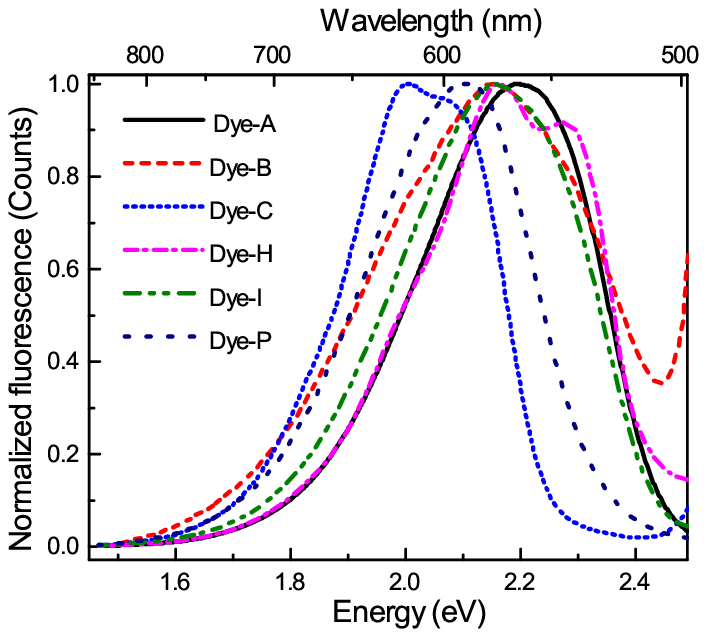}\includegraphics[width = 2.35 in]{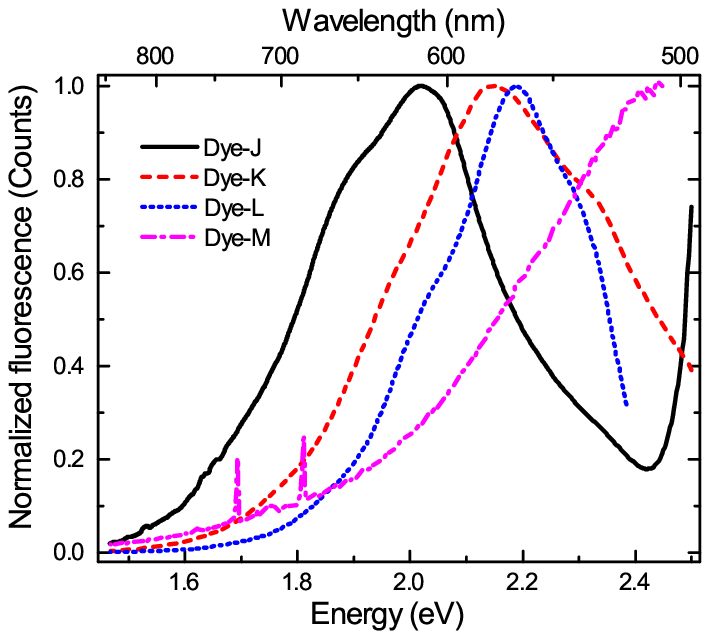}\includegraphics[width = 2.35 in]{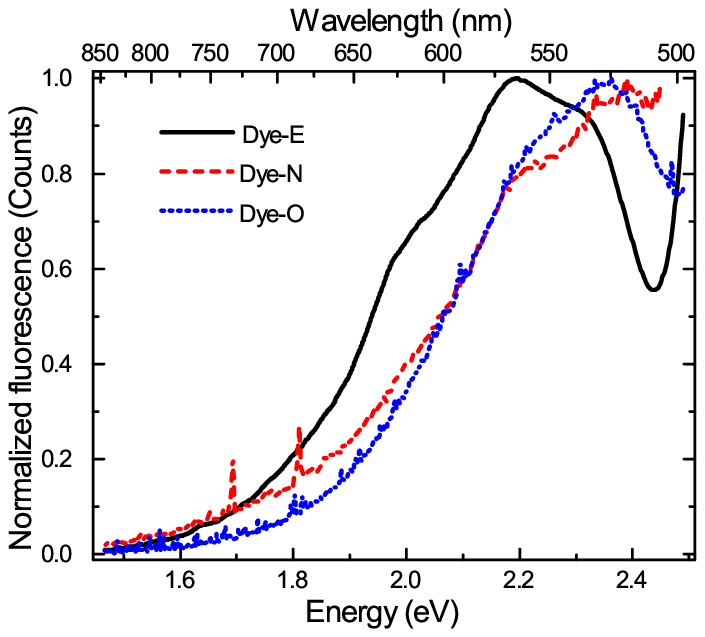}
\caption{Absorbance and fluorescence spectra for the amine class of dyes (left), hydroxy class (middle) and others (right).\label{fig:flu-amine}}
\end{figure*}

Samples for fluorescence experiments are made with a thermal press method, as follows.  Methyl methacrylate (MMA) monomer, a liquid, is run through a column to remove the inhibitor and filtered to remove particulates.  The AQ derivative is then dissolved into the monomer, using 45 to 60 minutes of sonication to ensure that the dyes are fully dissolved.  Butanethiol, a polymerization initiator, and tert-butyl peroxide, a chain transfer agent, are added to initiate polymerization and control polymer chain length.  The solution is sonicated for 10 to 15 minutes and then passed through a filter with $0.2 \mu m$ pore size to remove remaining particulate or dye aggregates.

The solution is then heated to $95^o C$ for about 50 hours to polymerize it.  After the hardened material is cooled and removed from the bottle, it is crushed to make small chips, which are placed into a custom oven at $150^o C$, where it is squeezed between two glass plates for 90 minutes at a pressure of 110 to 120 psi to form a thin film of $250$ to $300\,\mu m$ thickness.  All samples are prepared to concentrations of 3 grams of dye to 1 liter of MMA monomer.

\section{Results and Discussions}

Figure \ref{fig:flu-amine} shows the absorption and fluorescence spectra of the the three dye classes. All spectra are normalized to the largest value so that the peaks can be viewed together.  In the class designated ``other," the absorption peaks are outside the visible range in the UV end of the spectrum.  For those dyes that fluoresce, the peaks in the other class are above 2.2 eV.  In contrast, the amine and hydroxy classes show fluorescence peaks at about or below 2.2 eV.  A survey of the structures shows that absorption peaks are associated with the presence of a hydroxy or amine group adjacent to the central oxygen.  These are also associated with molecules that self heal, suggesting that the oxygen and adjacent group are together responsible.

\begin{figure}
\begin{center}
\includegraphics[width = 3.25 in]{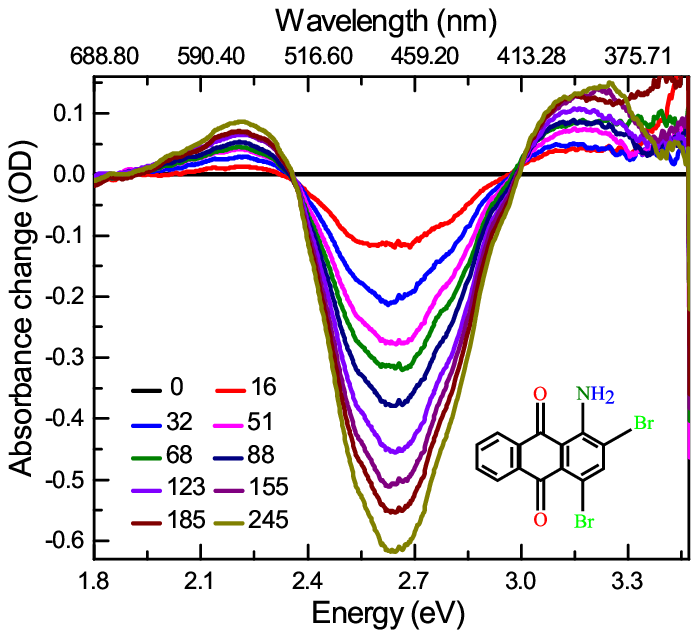}
\includegraphics[width = 3.25 in]{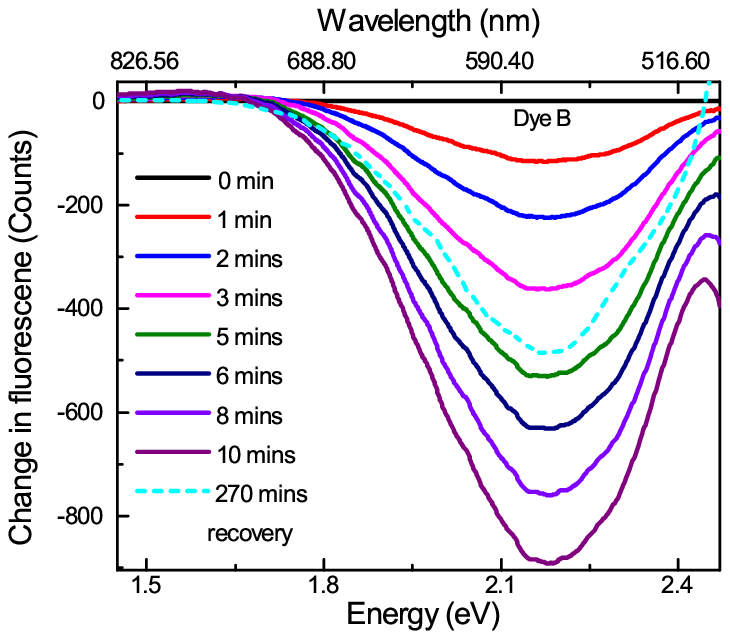}
\includegraphics[width = 3.25 in]{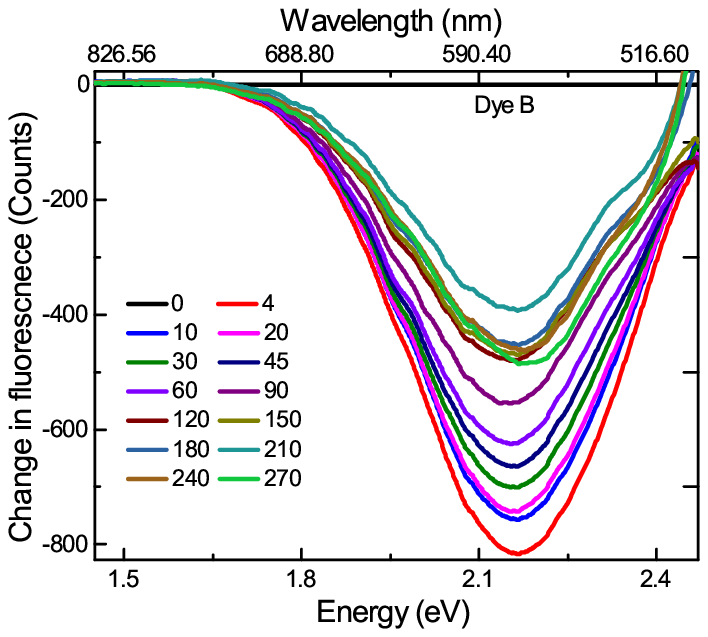}
\end{center}
\caption{Change in absorbance spectrum (top); and, fluoresce spectrum (center) of Dye B during photodegradation and (bottom) recovery.  The light dashed blue curve on the decay plot shows the spectrum after the sample has recovered for the time specified.}
\label{Fig:Abs-Flu change}
\end{figure}

The top portion of Figure \ref{Fig:Abs-Flu change} shows the change in absorption spectrum relative to a fresh sample of Dye B.  The isosbestic points at 2.36 eV and at 3.00 eV are indicators that one species (the fresh material) is being converted to another one -- the reversibly degraded species.

An isosbestic point can be understood as follows.  If the spectrum of the fresh material is given by the function $f(E)$ and the degraded species by $g(E)$, and the destruction of one fresh molecule leads to the creation of one degraded molecule, the difference spectra plotted in the top portion of Figure \ref{Fig:Abs-Flu change} is given by
\begin{equation}\label{Eq:Isosbestic}
[(1-a)f(E) + a g(E)] - f(E) = a(g(E)-f(E)),
\end{equation}
where $a$ is the fraction of molecules that have been damaged.  Equation \ref{Eq:Isosbestic} shows that where two spectra intersect, the difference spectrum will always vanish at that point, independent of the amount of damaged species produced.  If the spectrum of the damaged species were peaked far away from the fresh material's peak, the energy where the two intersect would be in a spectral region where the absorbance is low, so no isosbestic point would be observed.  In this case, the absorbance at the peak would be directly proportional to the population of undamaged molecules.

The data for Dye B suggests that both species are contributing to the absorbance near the peak of the undamaged molecule, so the change in strength of absorption alone cannot be used to determine the degree of degradation unless the absorption spectrum of the damaged species can be separately determined and subtracted from the measured absorbance.  Such a determination is difficult given the fact that the reversibly recovering species is healing, so its concentration cannot be unambiguously deduced.  As such, linear absorption spectroscopy as a tool for determining the populations suffers from a degree of uncertainly that cannot be eliminated.

For Dye B, the shape of the difference spectrum and its two isosbestic points that straddle the dip corresponding to a peak in the absorption maximum of the undamaged sample suggests one of two possibilities; the absorption spectrum of the damaged species peaks at approximately the same energy as the undamaged species but is of broader width, or, the damaged species has two peaks -- one on either side of the absorption peak of the undamaged species.  In contrast, the fluorescence spectrum in the bottom portion of Figure \ref{Fig:Abs-Flu change} shows only weak evidence of an isosbestic point at 1.57 eV.  {\color{red} Note that the fluorescence spectra during recovery also start at $t=0$, so the time during decay is the time after the pump beam is turned on and for recovery is the time after the pump beam is turned off.  These time origins are chosen for the purpose of data fitting.  All decay and recovery plots follow this convention.}

Since the linear absorption spectrum shows isosbestic points, and the fluorescence spectrum should appear as the approximate mirror image of the absorption spectrum, similar isosbestic points should be observed in the difference fluorescence spectra as in the difference absorption spectra.   The observed Stokes shift for Dye B between the absorption peak (2.64 eV) and fluorescence peak (2.15 eV) is 0.49 eV.  Using the fact that the isosbestic points in the difference absorption spectra are at -0.28 eV and +0.26 eV from the peak, applying the mirror image rule to the Stokes shift would imply that the fluorescence isosbestic points should be located at 1.89 eV and 2.43 eV.

There is no evidence of an isosbestic point at 2.43 eV, though this value is so close to the 2.54 eV pump laser that its presence could be masked.  On the other hand, there is no evidence of an isosbestic point at 1.92 eV, and what may be an isosbestic point at 1.57 eV is too far away in energy to be associated with the predicted placement.  The fluorescence spectrum for Dye B is consistent with a damaged species with disallowed or suppressed fluorescence from its excited state.  Indeed, if the fluorescence yield from the damaged species is smaller than from the fresh molecule, the isosbestic point would be shifted from the predicted value.  The observed isosbestic point is consistent with a highly suppressed fluorescence signal from the damaged species.

\begin{figure}[h!]
\begin{center}
\includegraphics[width = 3.25 in]{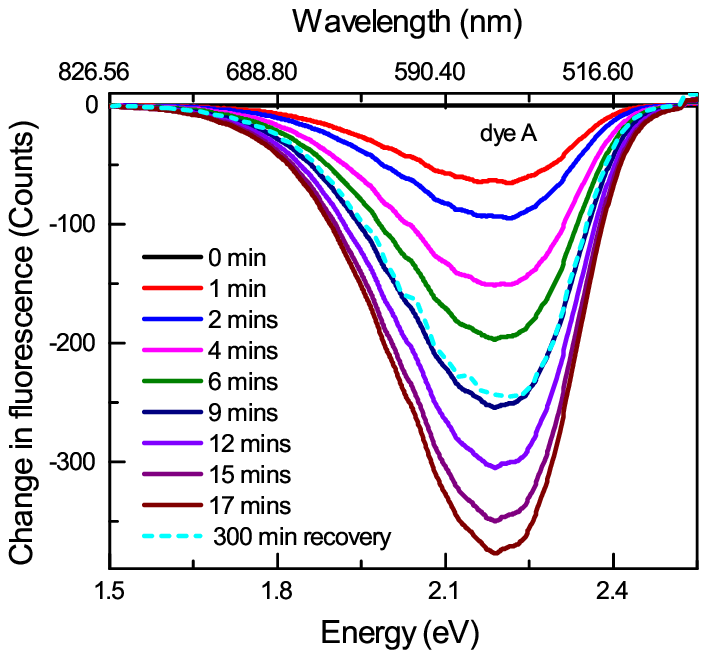}
\includegraphics[width = 3.25 in]{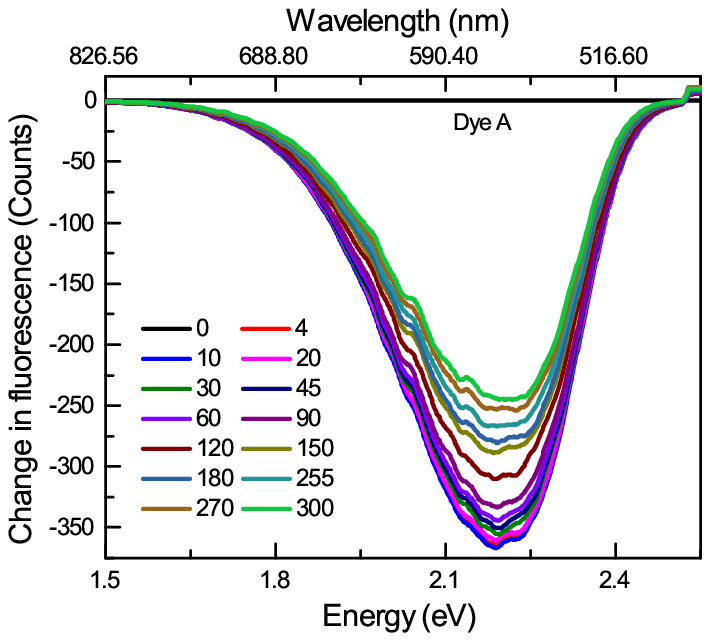}
\end{center}
\caption{Change in the fluorescence spectra of Dye A during decay (top) and recovery (bottom).  The light dashed blue curve on the decay plot shows the spectrum after the sample has recovered for the time specified.}
\label{fig:DyeA-Flu}
\end{figure}
Figure \ref{fig:DyeA-Flu} shows the change in fluoresce as a function of time during degradation of Dye A, also known as DO11.  The fluorescence spectra show no evidence of isosbestic points, though the change in the linear absorption spectra show distinct isosbestic points at 2.3 eV and 2.9 eV.\cite{embay08.01} This observation is consistent with reports that amplified spontaneous emission from the damaged species of the disperse orange 11 chromophore (Dye A) is undetectable compared with the strong signal recorded for the fresh dye.\cite{howel02.01,embay07.01,ramin11.01,ramin12.01} These results for Dye A and Dye B suggest that the undamaged species of these two anthraquinone dyes fluoresce while the damaged species do not, or at least the fluorescence signal of the damaged species is negligibly small.  With linear fluorescence, then, it is reasonable to assume that the measured fluorescence intensity is proportional to the population of undamaged molecules.

Two factors need to be taken into account to relate the measured fluorescence intensity to the population of undamaged molecules.  First, the fluorescence light generated inside the sample is absorbed by the intervening material on its way out.  If the absorbance of the sample is also changing upon damage, the measured change of fluorescence intensity is a convolution of the change in population and the change in the amount of light absorbed.  Secondly, since the sample absorbs the pump light, the intensity decreases as a function of depth, so the damaged population near the surface is greater than deeper within, where the pump intensity is lower.  As such, even if the fluorescence light was to be unattenuated, the measured signal would be proportional to the average population over the attenuation depth of the pump.  However, changes in the population of the damaged species can also change the amount of absorption, leading to additional complications.  So for example, if the absorption cross section of the damaged species is lower than the starting molecule, as it usually is at the pump wavelength (488 nm), the pump beam will penetrate further into the sample as that material near the surface is damaged.  As such, the less-damaged material inside will be probed, underestimating the degree of damage.

The effects of time-dependent absorption of the fluorescence light can be removed simply by using the fluorescence intensity at the isosbestic point as a probe.  This requires a preliminary determination of the isosbestic point from the absorption spectrum as a function of time during photodegradation.  For a thin enough sample, the effect of pump depletion with depth is negligible.  However, generating a suitably measurable fluorescence intensity often requires thicker samples, so the tradeoffs between pump depletion and fluorescence signal strength must be carefully considered.

\begin{table}\caption{Sample and molecule properties at the pump wavelength of $488 \, nm$.  Check marks indicate samples that heal.}
\begin{center}
    \begin{tabular}{  l  p{.8in}  p{.98in} p{.85in}}
    \hline
    Dye & Molecular Number Density  & Absorption Cross Section & 1/e Absorption Length\\
    & $(\times 10^{18} cm^{-3})$ & $(\times 10^{-18} cm^2)$ & $(\mu m)$ \\
    \hline\hline
    A \checkmark ~~~ & 7.6 &  16 & 81 \\
    B \checkmark & 4.7 &  28 & 76 \\
    C & 7.6 & 2.7 & 490 \\
    \rowcolor{red} D & 8.7 & 1.1 & 1000 \\
    E \checkmark & 7.0 & 9.3 & 150 \\
    \rowcolor{red} F & 7.5 & 0.83 & 1600 \\
    \rowcolor{red} G & 7.5 & 0.41 & 3300 \\
    H \checkmark & 7.6 & 20 & 67 \\
    I \checkmark & 7.6 & 18 & 72 \\
    J \checkmark & 7.5 & 17 & 78 \\
    K \checkmark & 7.5 & 29 & 46 \\
    L \checkmark & 7.5 & 9.2 & 140 \\
    M & 7.5 & 4.2 & 320 \\
    \rowcolor{red} N & 7.6 & 1.4 & 930 \\
    O & 8.1 & 1.8 & 690 \\
    P \checkmark & 7.6 & 7.6 & 170 \\
    \hline
    \end{tabular}
\end{center}\label{tab:CorssSections}
\end{table}
In the present experiments, most of the samples are thicker (between $250 \mu m$ and $300 \mu m$) than the peak $1/e$ absorption length.  Table \ref{tab:CorssSections} shows the absorption cross sections and $1/e$ absorption lengths at the absorption maximum for the molecules studied.  If the pump intensity profile is to remain constant, the change in absorbance during photodegradation should be small relative to the initial absorbance.  We focus on dye B as an example of the tradeoffs.

\begin{table}\caption{Dye B properties at key wavelengths.}
\begin{center}
    \begin{tabular}{ p{.75in}  p{.65in}  p{.9in} p{.8in}}
    \hline
    Wavelength (Energy) & Comment   & Absorption Cross Section & 1/e Absorption Length\\
    nm (eV)& & $(\times 10^{-18} cm^2)$ & $(\mu m)$ \\
    \hline\hline
    488 (2.54) & undamaged at pump &  ~~~~ 24 & ~~~~ 88 \\
    488 (2.54) & damaged at pump &  ~~~~ 16 & ~~~~ 140 \\
    575 (2.54) & undamaged at fluorescence peak & ~~~~ 1.2 & ~~~~ 1800 \\
    575 (2.54) & damaged at fluorescence peak & ~~~~ 2.4 & ~~~~ 870 \\
    526 (2.36) & isosbestic point \#1 &  ~~~~ 54 & ~~~~ 390 \\
    414 (2.99) & isosbestic point \#2 &  ~~~~ 12 & ~~~~ 170 \\
    \hline
    \end{tabular}
\end{center}\label{tab:CorssSectionsKey}
\end{table}

Table \ref{tab:CorssSectionsKey} shows a summary of the absorption cross-sections and absorption lengths for the sample made from Dye B.  The $1/e$ absorption length at the pump wavelength is about $88 \mu m$, which is much shorter than the typical sample thicknesses of $250-300 \, \mu m$.  After maximum burn, the $1/e$ absorption length increases to $140 \, \mu m$.  As such, the concentration profile of damaged molecules spreads deeper into the sample; but, most of the pump light is absorbed within the sample for the duration of the burn.  As a result, the same number of photons are absorbed.  Since the damaged molecules do not fluoresce, the fluorescence power is reduced in proportion to the number of damaged molecules present that absorb the pump light.

If the sample were transparent at the fluorescence measurement wavelength, changes in the measured fluorescence intensity would be be an accurate representation of the fractional change of the number of pristine molecules due to decay or recovery.  Because light from the deeper regions must travel further before exiting, more of the fluorescence is absorbed if the optical density of the sample increases at the measurement wavelength.

For a ballpark estimate of the amount of fluorescence light that is absorbed on its way to the detector, we assume that the light is on average produced at a point in the sample where the pump intensity is attenuated by a factor of $1/e$, which for the undamaged sample is $88 \mu m$.  The $1/e$ absorption length of the light emitted at the observed fluorescence peak is $1,800 \mu m$.  As such, the fraction of fluorescence signal that makes it out of the sample from the interior is given by $\exp[-88/1800] = 0.95$.  At the end of the burn experiment, the pump light penetrates $140 \mu m$ into the sample, at which time the $1/e$ absorption length at the fluorescence peak is $870 \mu m$.  The fraction of the light making it to the detector is then given by $\exp[-140/870] = 0.85$.  As such, the change in fluorescence signal observed during an experiment will be due to a convolution of: (1) the change in the amount of damaged molecules produced -- which do not fluoresce; (2) the longer distance that the light needs to travel to exit the sample; and, (3) the larger optical extinction of the damaged molecules that the light must traverse.  The latter two effects combine to decrease the fluorescence signal by about 10\%.  Since the observed fluorescence is a linear process, this self-absorption effect leads to an underestimate of the undamaged population by about 10\%.

\begin{figure}[h!]
\begin{center}
\includegraphics[width = 3.25 in]{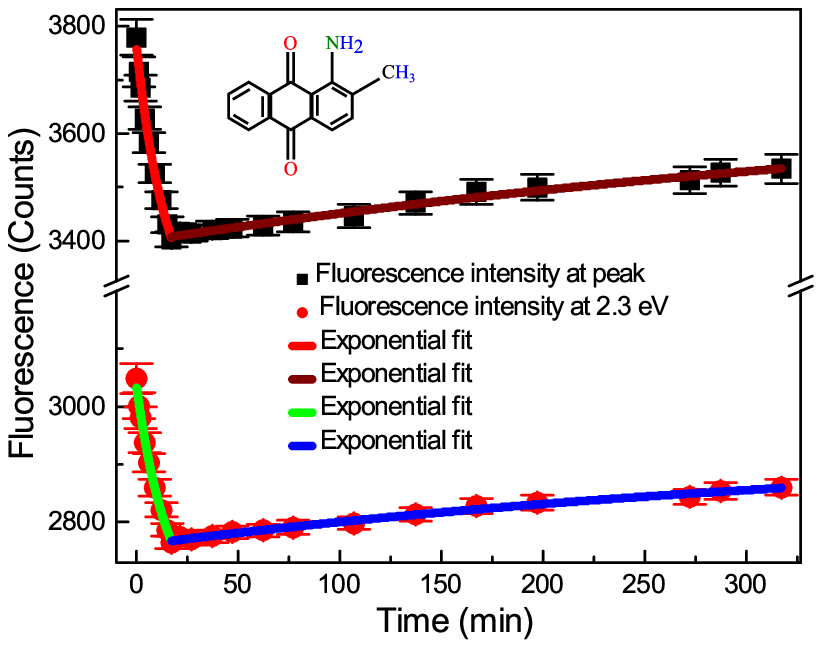}
\includegraphics[width = 3.25 in]{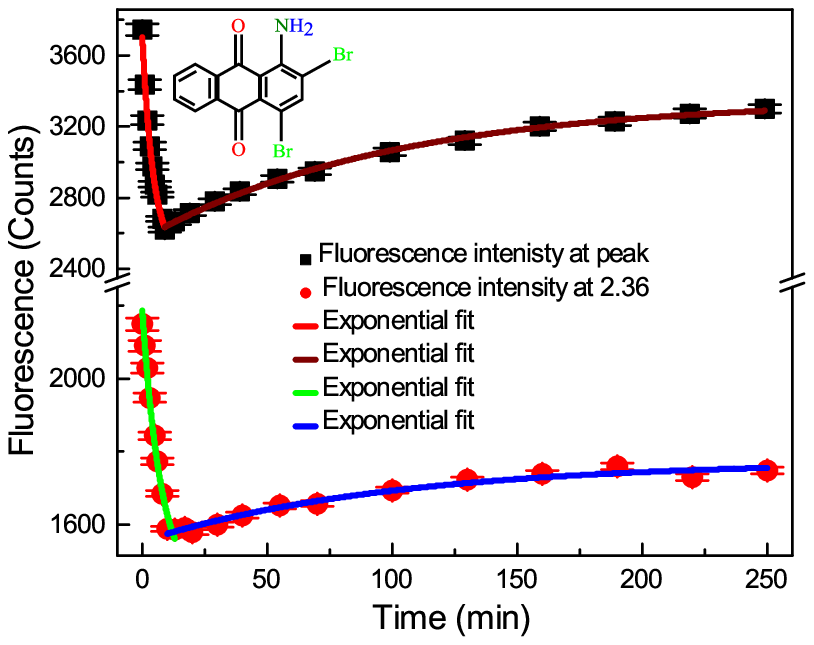}
\end{center}
\caption{Fluorescence intensity of Dye A (top) and Dye B (bottom) as a function of time at the fluorescence peak and at the isosbestic point.  The points are the data and the curves are exponential fits.}
\label{fig:Flu-dec-rec-peak-iso}
\end{figure}

A strategy for avoiding problems associated with the change in absorbance of the sample during burning is to tune the pump laser to one of the isosbestic points - one of which falls conveniently near the $488 \, nm$ pump line; and to probe fluorescence at the other isosbestic point.  It is far easier to isolate one part of a fluorescence spectrum than to tune the pump laser, so we concentrate on evaluating the approach of fluorescence probing at the isosbestic point.

To test this strategy, we first consider the degree of degradation determined from the fluorescence power at the peak and at the isosbestic point during photodegradation.  The data in Figure \ref{fig:Flu-dec-rec-peak-iso} shows that at the peak, the fluorescence power falls to $0.70 \pm 0.01$ of its initial value during a degradation experiment -- a 30\% decrease.   At the isosbestic point, in the same experiment and at the same time, the intensity falls to $0.74 \pm 0.01$ of its initial value -- a 26\% decrease.  Clearly, both experiments give values that are close but outside of experimental uncertainties. However, a back-of-the envelope calculation of self absorption above predicts that the measurement at the peak should result in a 10\% larger decrease in the fluorescence intensity than at the isosbestic point. Thus, given the actual decrease of 26\% at the isosbestic point, the self-absorption calculation predicts that at the fluorescence peak, the decrease should be 29\%, in good agreement within experimental uncertainties.

\begin{figure}[h!]
\begin{center}
\includegraphics[width = 3.25 in]{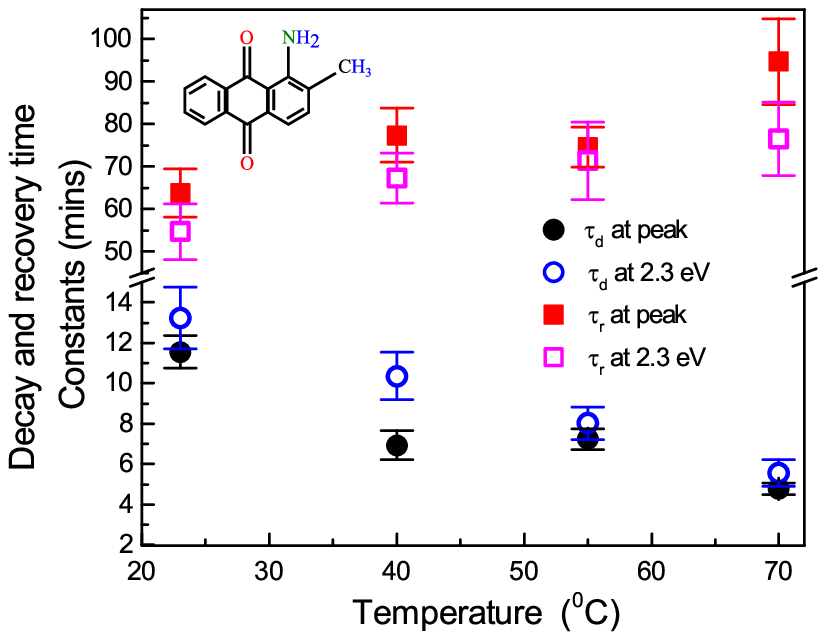}
\includegraphics[width = 3.25 in]{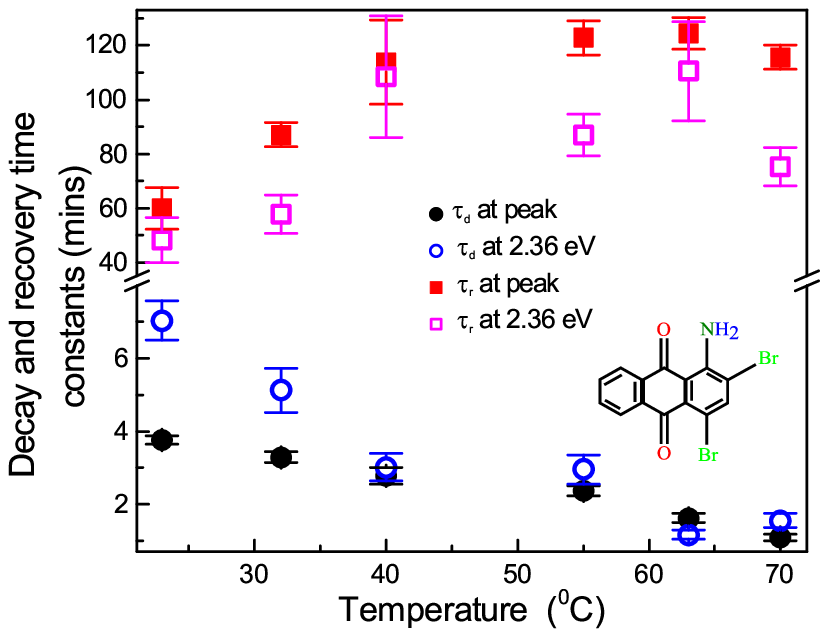}
\end{center}
\caption{Time constants of Dye A (top) and Dye B (bottom) as a function of temperature determined from the fluorescence peak and at the isosbestic point.  $\tau_d$ and $\tau_r$ are the decay and recovery time constants, respectively.}
\label{fig:Flu-time_constants-B}
\end{figure}

Figure \ref{fig:Flu-dec-rec-peak-iso} shows the fluorescence intensity at the isosbestic point and at the peak for Dyes A and B during decay and recovery at room temperature.  The curves are fits to an exponential.  The same experiment is repeated as a function of temperature and the time constants determined from the fits are plotted as a function of temperature in Figure \ref{fig:Flu-time_constants-B}.  The decay and recovery time constants determined from probing at the isosbestic point are for the most part within experimental uncertainties of the values determined from the peak fluorescence.  However, the decay time constants determined from the peak fluorescence intensity are systematically below the ones determined at the isosbestic points while the recovery time constants tend to be larger when determined at the peak.

Though an accurate determination of the time constants requires that the fluorescence probe wavelength be coincident with the isosbestic point to avoid changes in self absorption that bias the data, time constants determined from data taken at the fluorescence peak are reliable enough for studies that seek to compare trends in different chromatophores.  As such, absorption measurements are generally not needed to determine the isosbestic points unless accurate numerical results are needed.  We use the approximate approach to study a series of anthraquinone dyes to search for structural features in molecules that are associated with healing and to determine if the temperature dependence of recovery follows a barrier model or a domain model.  These will be described later.

\begin{figure*}
\includegraphics[width = 2.35 in]{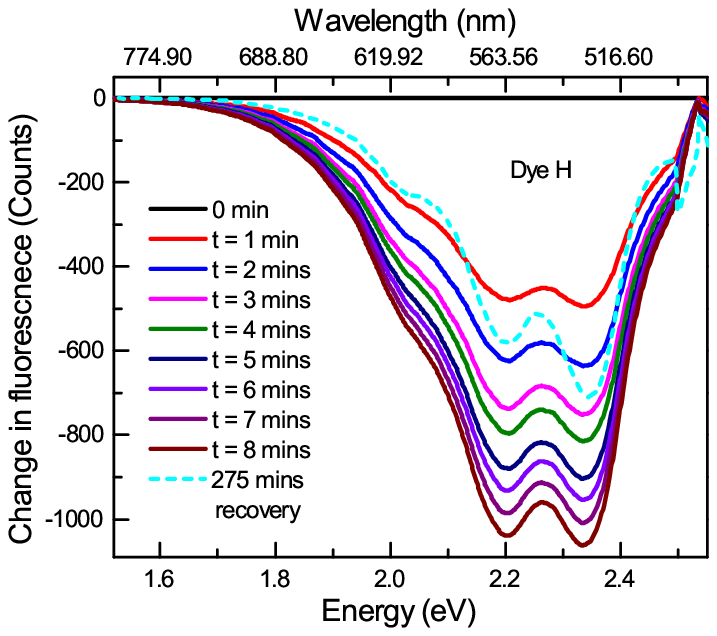}\includegraphics[width = 2.35 in]{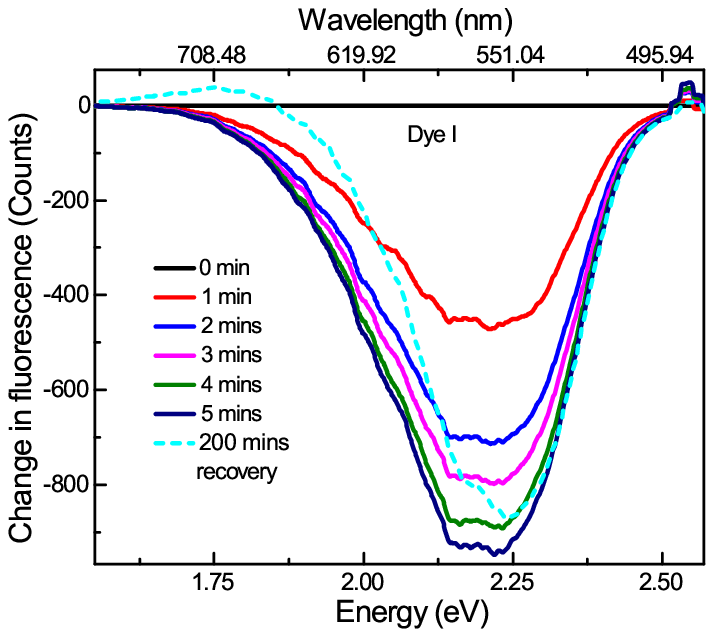}\includegraphics[width = 2.35 in]{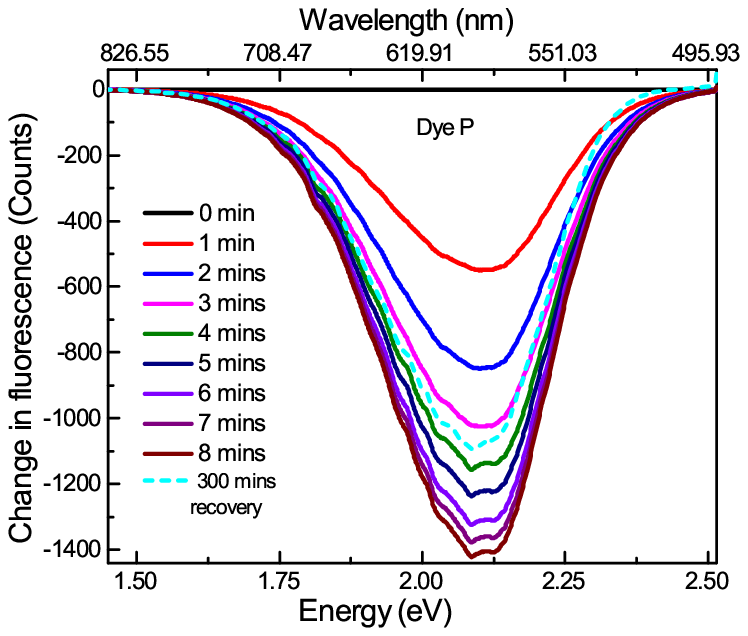}
\includegraphics[width = 2.35 in]{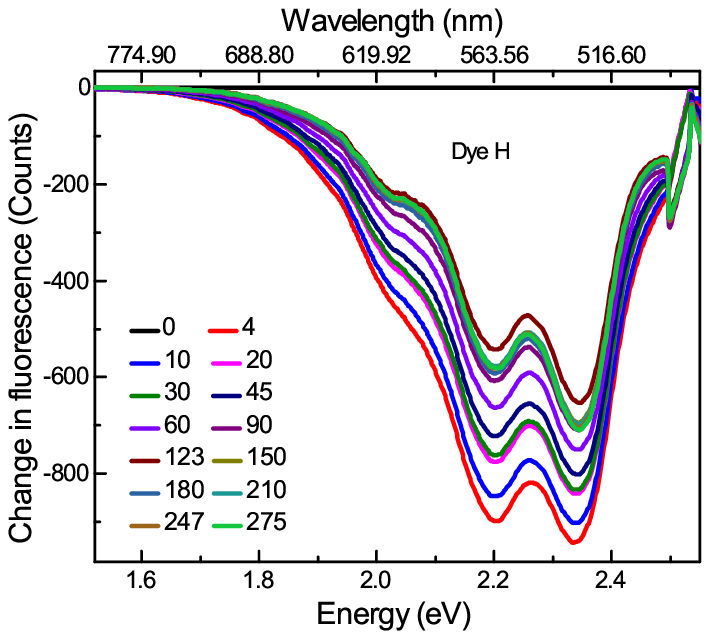}\includegraphics[width = 2.35 in]{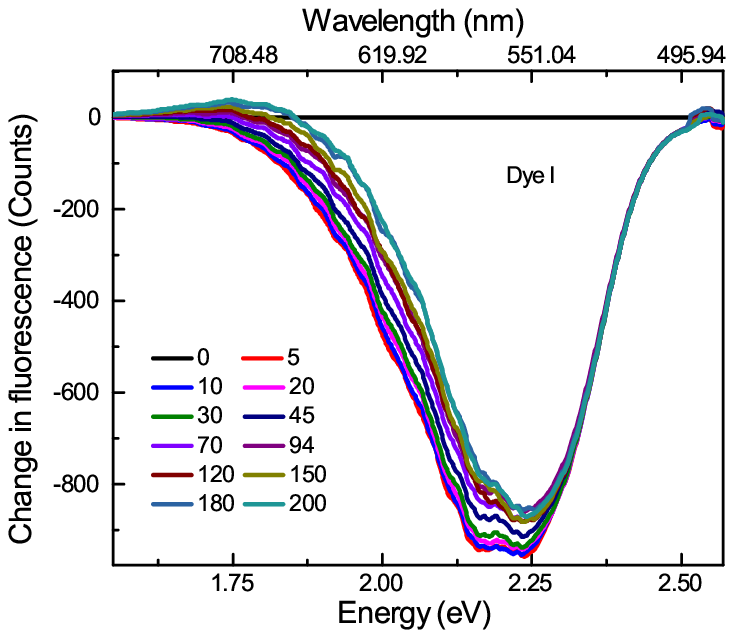}\includegraphics[width = 2.35 in]{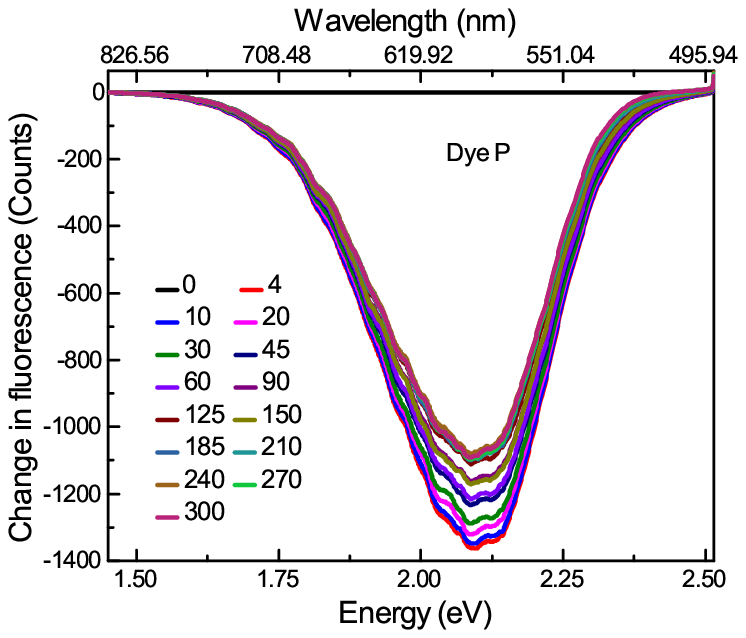}
\caption{Change in the fluorescence spectrum of the amine class of molecules during decay (top row) and recovery (bottom row) of Dye H (left column), Dye I (middle column) and Dye P (right colum).  The light dashed blue curves on the decay plots show the spectrum after the samples have recovered for the time specified.\label{fig:flu-dec-rec-amines}}
\end{figure*}

\begin{figure*}
\includegraphics[width = 2.35 in]{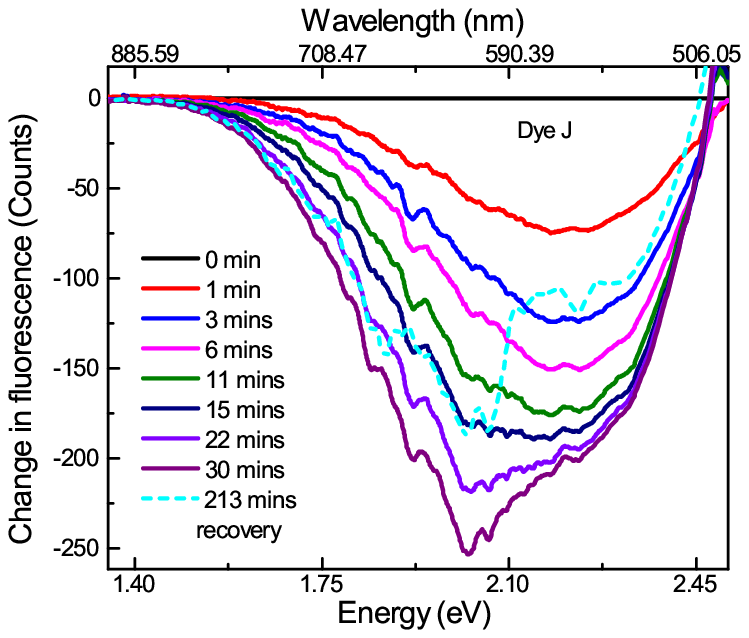}\includegraphics[width = 2.35 in]{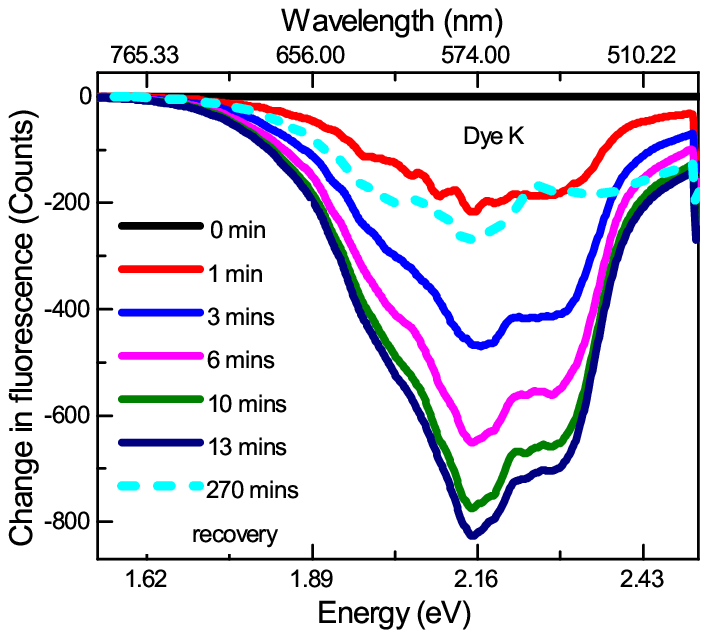}\includegraphics[width = 2.35 in]{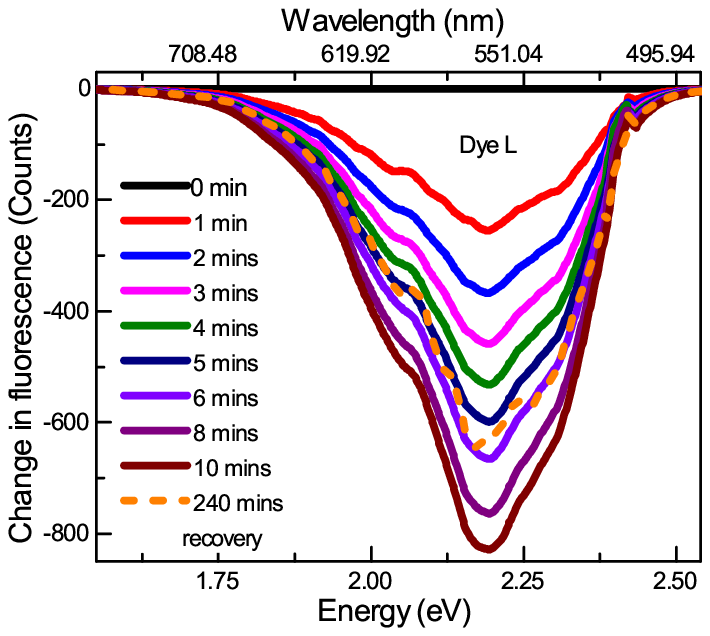}
\includegraphics[width = 2.35 in]{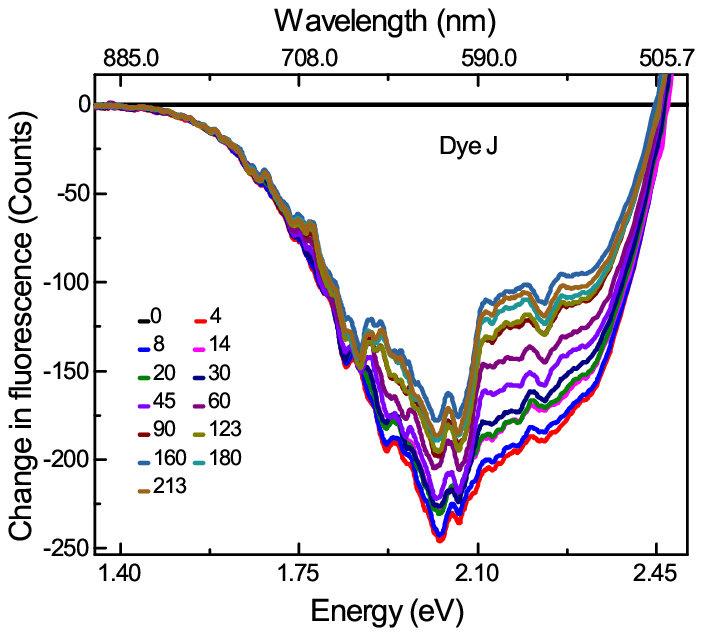}\includegraphics[width = 2.35 in]{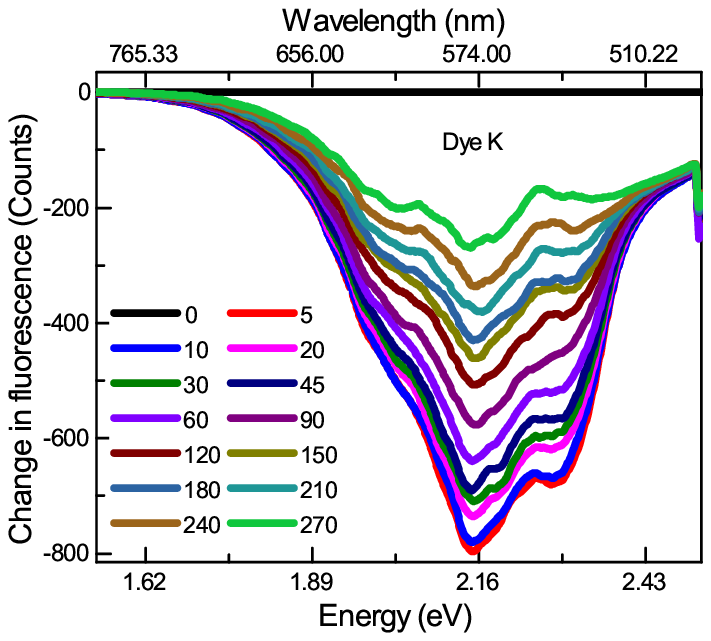}\includegraphics[width = 2.35 in]{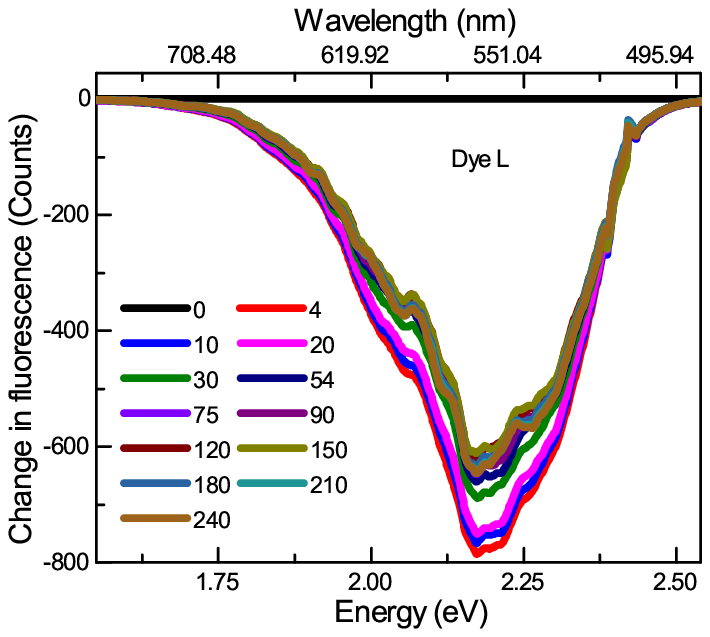}
\caption{Change in the fluorescence spectrum of the hydroxy class of molecules during decay (top row) and recovery (bottom row) of Dye J (left column), Dye K (middle column) and Dye L (right column). The light dashed blue curves on the decay plots show the spectrum after the samples have recovered for the time specified.\label{fig:flu-dec-rec-hydroxy}}
\end{figure*}

Figures \ref{fig:flu-dec-rec-amines} and \ref{fig:flu-dec-rec-hydroxy} show the change in the fluorescence spectra at room temperature during photodegradation and recovery for the amine and hydroxy group of dyes.  The light blue dashed curve shows the spectrum after recovery superimposed on the sets of curves taken during decay for comparison.  All the spectra have in common several features.  First, the recovery times are much longer than photodegradation times and the fluorescence spectrum never fully recovers over experimental timescales.  The implication is that either an irreversibly damaged species is formed that never recovers, or there are at least two mechanisms for healing, in which case one of the mechanisms acts over timescales that exceed typical experimental durations.

The light blue dashed curves that are superimposed on the fluorescence spectra taken during decay are generally of approximately the same shape as the spectra during photodegradation. Small differences can be attributed to drift in the spectrometer and the apparatus.  These data support the hypothesis that one mechanism is at work.  Alternatively, if two species are being formed -- and only one recovers, the fact that the recovery spectra retrace the decay spectra in reverse would suggest that the two species have the same spectral shape within experimental uncertainties.  The third and most likely hypothesis is that all of the degraded species, reversible or not, do not fluoresce, so only the undamaged species is probed.  One notable exception is Dye J, where it appears that an irreversibly-damaged species is produced during photodegradation that does not recover but produces a feature in the fluorescence signal.  The purported irreversible species is associated with the dip at $2.0 \,eV$, which does not recover, and is superimposed on the broader fluorescence peak, just above $2.2 \, eV$, of the recovering species.  Most likely, this feature is due to the formation of a species that absorbs the fluorescence light.  As such, the decrease in fluorescence is most likely due to the combined effect of the depletion of undamaged molecules, which is responsible for the dip at above $2.2 \, eV$ and the accumulation of an irreversibly-damaged species that absorbs light at $2.0 \, eV$

\begin{figure}[h!]
\begin{center}
\includegraphics[width = 3.25 in]{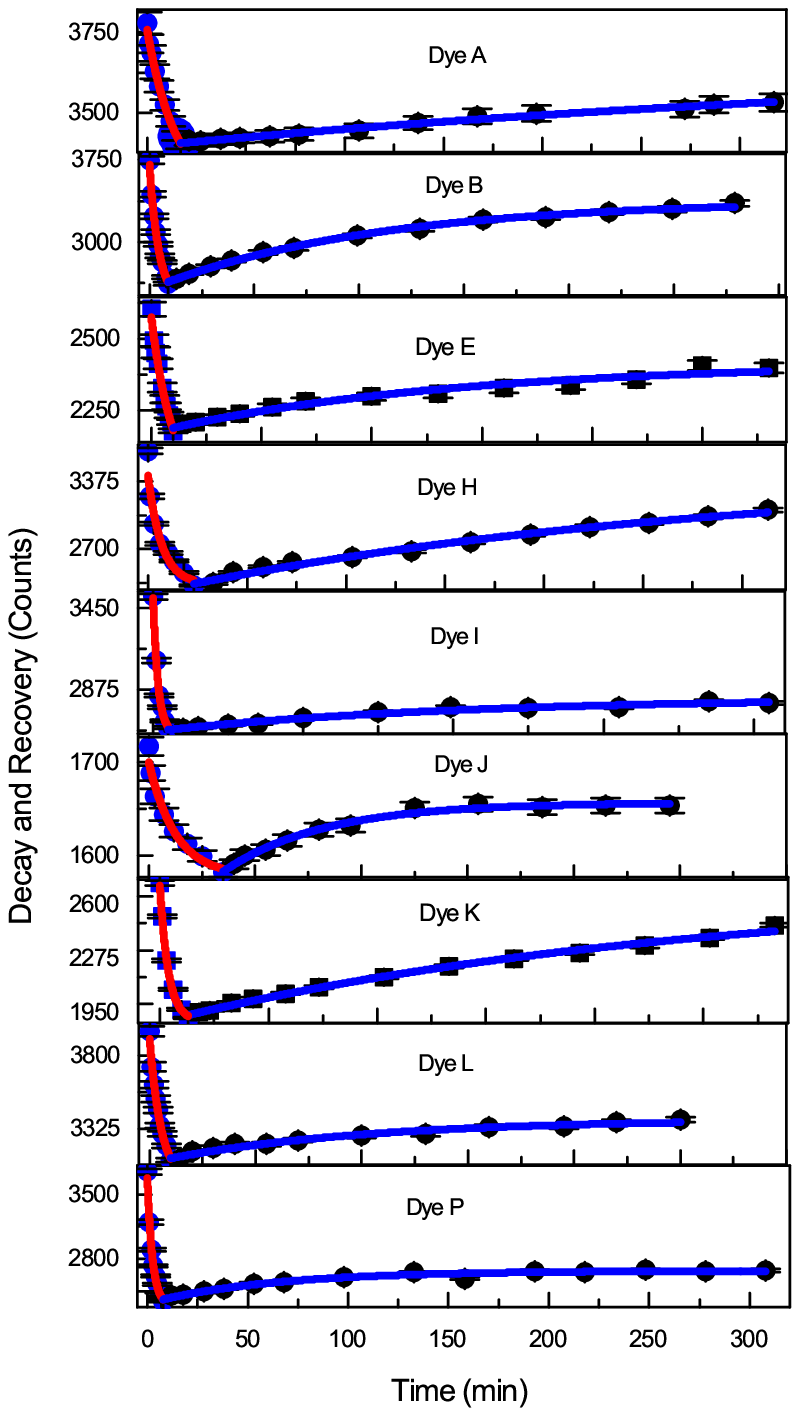}
\end{center}
\caption{Fluorescence as a function of time during decay and recovery at peak fluorescence of various AQ derivatives at room temperature.  The curves are fits to exponentials during decay (red) and recovery (blue).}
\label{fig:Flu-Dec-Rec}
\end{figure}

\begin{figure}
\centering
\includegraphics[width = 3.25 in]{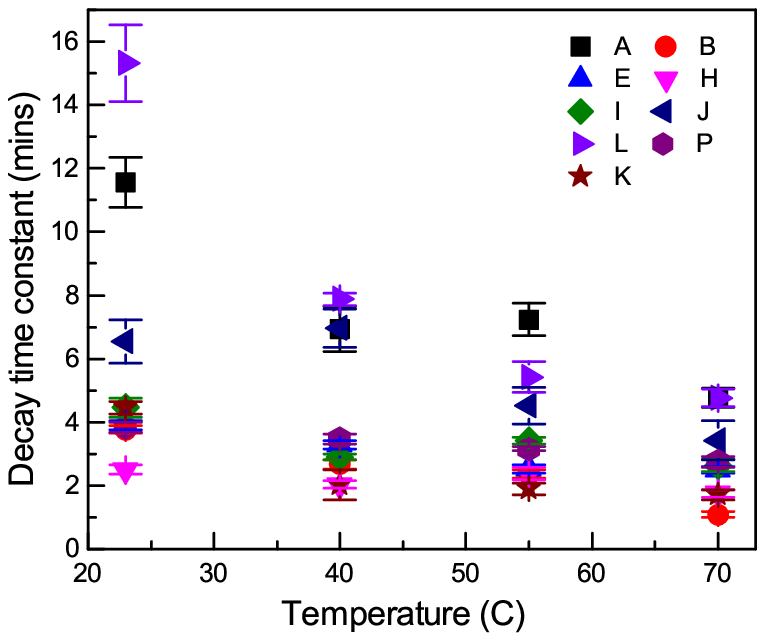} \includegraphics[width = 3.25 in]{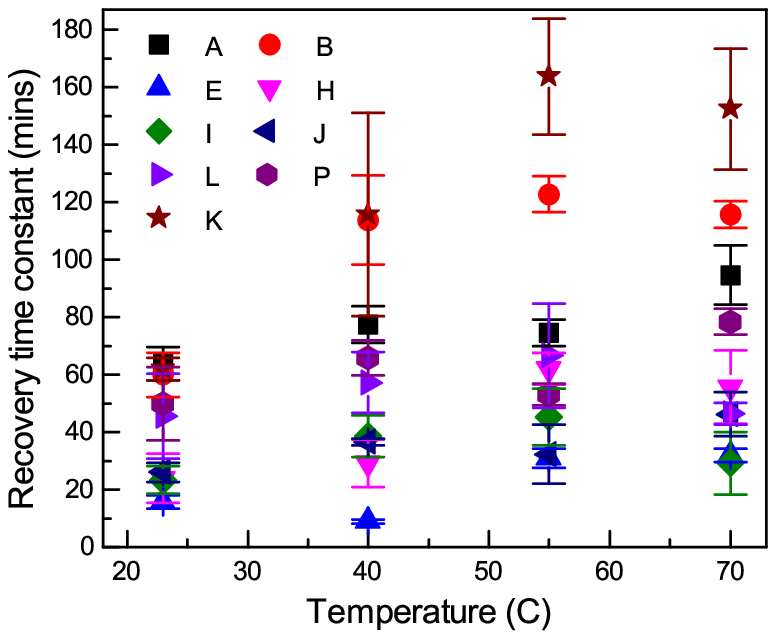}
\caption{Fluorescence decay time constant (top) and recovery time constant (bottom) of various AQ molecules as a function of temperature.}
\label{fig:Flu-Dec-Rec-Temp}
\end{figure}

Figure \ref{fig:Flu-Dec-Rec} shows the peak fluorescence intensity at room temperature as a function of time of the dyes that recover (points).  The curves are exponential fits to the data during decay (red) and recovery (blue).  The decay and recovery curves are measured and fitted to exponentials for several temperature to determine the decay and recovery time constants as a function of temperature. Figure \ref{fig:Flu-Dec-Rec-Temp} shows the results of such an analysis of all the dyes analyzed in Figure \ref{fig:Flu-Dec-Rec}.

There are two competing models of the self-healing mechanism, as described by Ramini and coworkers in each of two papers.\cite{ramin12.01,ramin13.01}.  The most intuitive one is that the photo-degraded species is of higher energy than the fresh molecule, but is trapped in a well with an energy barrier that prevents it from recovering.  In this scenario, an increase in temperature would increase the tunneling rate, and hence decrease the time constant.  The temperature dependence observed here and in past studies suggest that the time constant decreases with temperature.

The observation of deceased tunnelling rate with increased temperature in DO11 dye led Ramini and coworkers to propose the correlated chromophore domain mechanism, which proposed that molecules aggregate into protective domains, which foster healing in proportion to the size of the domain.\cite{ramin12.01,ramin13.01} As the temperature is increased, this model proposes that the competition between thermal jiggling and the chemical potential that binds molecules to a domain causes the domains to break apart.  The smaller domains then lead to a slower recovery rate.  This model fits the data well but is at odds with the predictions of a barrier model.  However, Hung has proposed that in similar dyes, these observations might be consistent with a Gaussian distribution of barrier energies, which yields a stretched exponential distribution of the time constants.\cite{hung15.01} The trends of the data here seem to contradict this hypothesis.

Another goal of the present work is to asses the structural features of a molecule that are associated with healing after photodegradation.  The trends in the data suggests that the key feature is the proximity of an $OH$ or $NH_2$ group to the central oxygen atom.  As a whole, proximity to the oxygen seems to be a common trait, which might be critical in understanding the mechanism.
\begin{figure}[h!]
\centering
\includegraphics[width = 3.25 in]{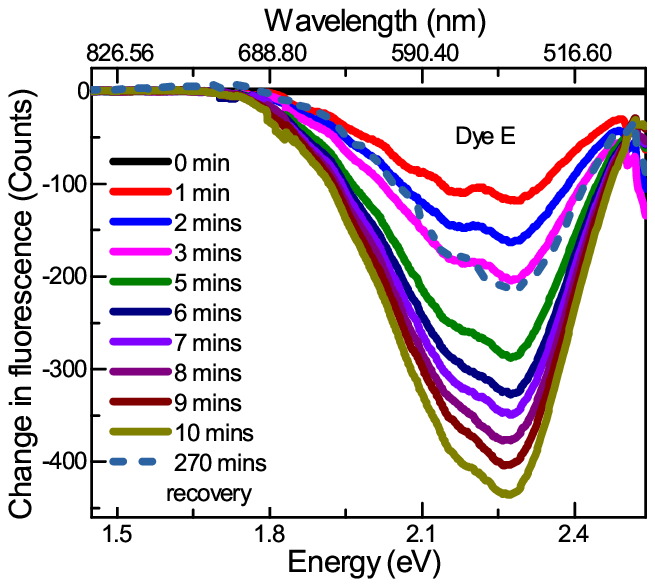}
\includegraphics[width = 3.25 in]{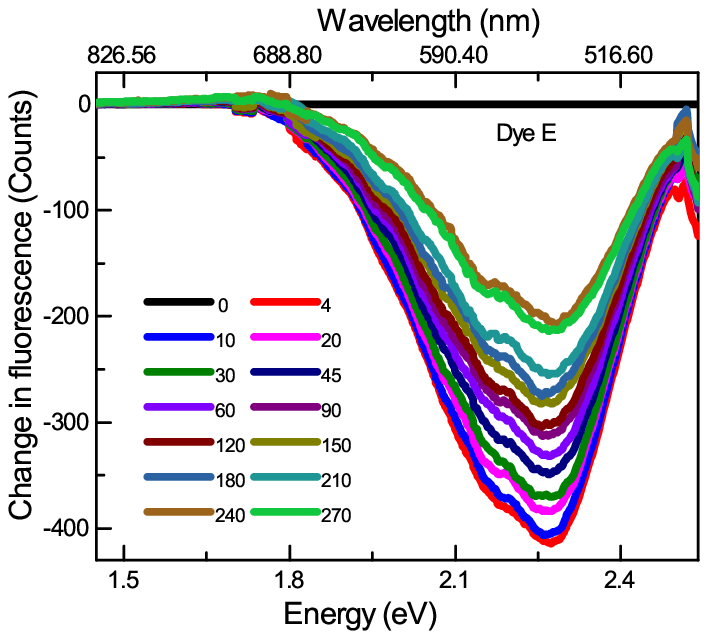}
\caption{Change in fluorescence spectrum as a function of time during decay (top) and recovery (bottom) of sample E.  The light dashed blue curve on the decay plot shows the spectrum after the sample has recovered for the time specified.}
\label{fig:false-color}
\end{figure}

Figure \ref{fig:false-color} shows the evolution of the fluorescence spectrum during decay and recovery of the fluorescence spectrum of Dye E.  The shape of the spectrum is the same during recovery, suggesting that the fresh material degrades into a species that recovers.  If multiple damaged species are formed, none of them emit fluorescence in the range measured.

Dye E does not fit well into the other two classes.  However, the ring to the upper left of the structure is near the central oxygen, thus following the pattern that proximity between the oxygen and a group attached to the adjacent site is the key feature for reversible photodegradation.

\begin{figure}[h!]
\centering
\includegraphics[width = 3.25 in]{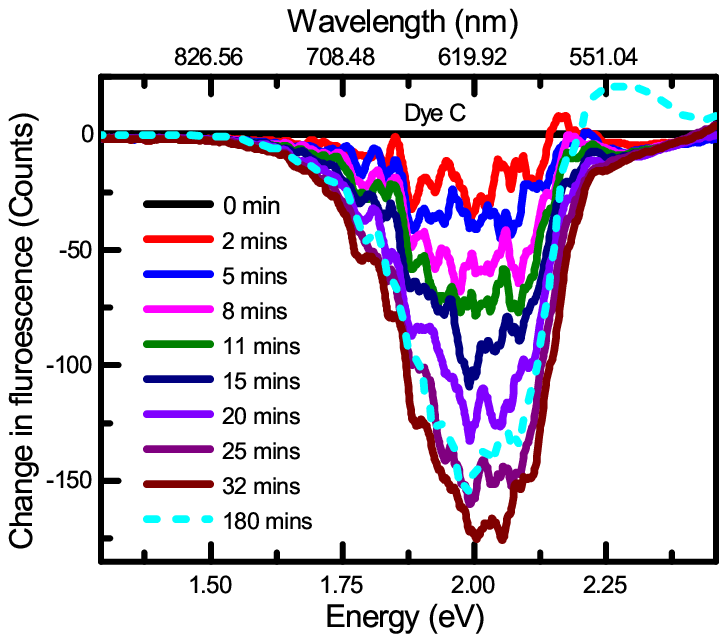}
\includegraphics[width = 3.25 in]{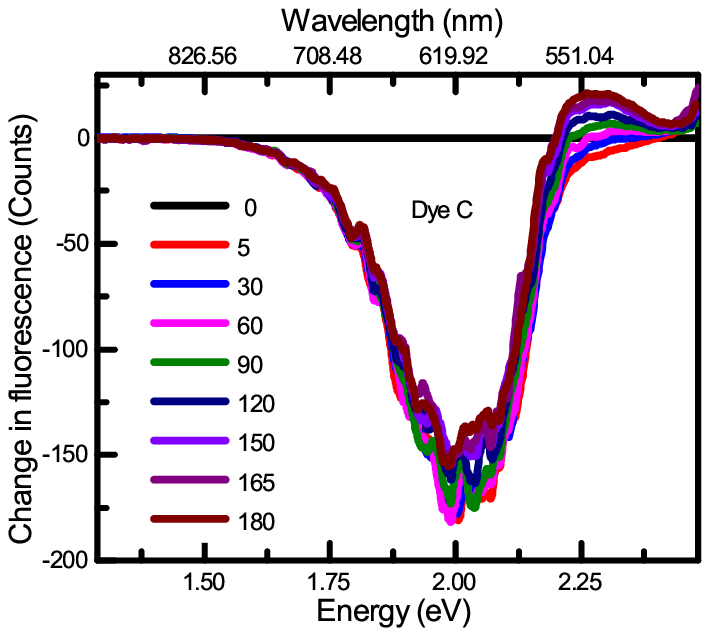}
\includegraphics[width = 3.25 in]{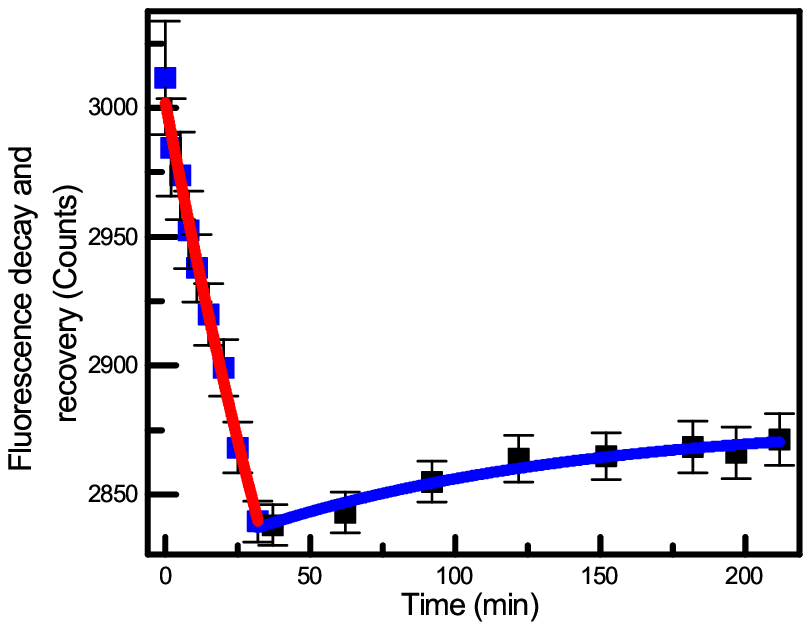}
\caption{\color{red}Change in fluorescence spectrum as a function of time during decay (top) and recovery (middle) of sample C in the one measurement that shows a small amount of recovery.  The light dashed blue curve on the decay plot shows the spectrum after the sample has recovered for the time specified.  The bottom graph shows the decay and recovery fluorescence signal at peak fluorescence with exponential fits.}
\label{fig:DyeC-Decay-Recover}
\end{figure}
{\color{red}Dye C is a problematic one because having one amine group and one hydroxy group, it is a combination of the two classes that show self-healing, yet recovery is ambiguous.  While the other molecules that recover do so reproducibly, Dye C is rarely observed to self-heal, and when it does, the degree of recovery is small.}

{\color{red} The top graph in Figure \ref{fig:DyeC-Decay-Recover} shows photodegradation in fluorescence.  The loss of fluorescence signal indicates photodegradation that is similar to the other dyes, so the material is clearly not damaged before the start of the experiment.  Furthermore, the linear absorption spectrum shows a well-defined peak characteristic of a fresh sample.  The middle graph in Figure \ref{fig:DyeC-Decay-Recover} shows one of the rare instances where recovery was observed, through the degree of recovery is small.  The bottom graph shows the signal at the peak during decay and recovery.}

{\color{red} Given that recovery is rare, and small when it is observed, we have characterized the material as non-recovering.  However, we admit that a small amount of recovery is possible.  It is puzzling that this molecule contains the groups associated with robust healing, yet does not significantly heal.  Perhaps this has something to do with the fact that it is a mixture of the two classes so that competing effects work against each other.  This observation calls for a more in-depth focus on this molecule, which is beyond the scope of the present paper, which seeks to find general trends associated with healing.}

\section{Conclusion}

Many questions remain.  Are the domains real physical structures or are they correlations between molecules that are mediated by coupling to a polymer chain?  Direct scattering measurements will need to be applied to search for aggregates of molecules distributed in size and space as predicted by the model, which would be strongly suggestive of real physical domains.  What are the nature of the domains and what are the forces that hold them together?  This question could be answered with Fourrier Transform Infra Red (FTIR) measurements that map out which bonds are affected during decay and recovery.

The present work on a set of varied molecules suggest that anthraquinone molecules with $NH_2$ and $OH$ groups adjacent to the central oxygen atom are the ones that self heal.  The importance of proximity of an atom or molecular group to the central oxygen is bolstered by the observation that Dye E, with a ring in that position, also heals. Furthermore, the fact that the observed recovery rates generally decrease with increased temperature suggests that the domain picture might the correct one.  However, more complex explanations that rely on distributions of barriers may revive activation energy models.  Fluorescence measurements, as described here, need to be applied to other systems that show self healing such as Rhodamine 6G\cite{ander15.02} to broaden our understanding of what types of structures are required for healing.

While each characterization method suffers from unique drawbacks, fluorescence in anthraquinone dyes seem to be immune from the need to correct for spectral changes inherent to the decay process because the degraded species appear to not fluoresce significantly where the emission wavelengths of pristine molecules peak.  Furthermore, the wavelength of the isosbestic point of the linear absorption spectrum can be used as the wavelength to monitor fluorescence, thereby eliminating the need to correct for the change in the absorbance of the probe light by the sample during photodegradation and recovery.  As such, the technique proposed here may be ideal for the future study of self healing.

Other refinements may be required to fully take advantage of this approach, including pumping at an isosbestic point to control for changes in the pump's penetration depth, and taking into account the fact that the pump intensity is brightest at the surface, where more of the irreversible species might be generated.  Though self healing has been observed in many materials, the mechanisms remain a mystery.  A deeper understanding of the processes at work will undoubtedly lead to interesting new science, and the fluorescence technique described here may be one of the best tools for the job.

\section{Acknowledgments}
We would like to thank Wright Patterson Air Force Base and Air Force Office of Scientific Research (FA9550- 10-1-0286) for supporting this research.


\end{document}